\documentclass[fleqn,usenatbib,useAMS]{mnras}

\usepackage{graphicx}
\usepackage{comment}
\usepackage{color}
\title[Time delays between radio and X-ray of Sgr A*]
 {Time delays between radio and X-ray and between narrow radio bands of Sgr A* flares in  the shock oscillation model }
\author[T. Okuda, C. B. Singh, R. Aktar]
 { Toru Okuda$^{1}$ \thanks{E-mail:bbnbh669@ybb.ne.jp},
  Chandra B. Singh$^{2}$\thanks{E-mail:chandrasingh@ynu.edu.cn},
  Ramiz Aktar$^{3}$\\
  $^{1}$ Hakodate Campus, Hokkaido University of Education, Hachiman-Cho 1-2, Hakodate 040-8567, Japan \\
  $^{2}$  South-Western Institute for Astronomy Research, Yunnan University, University Town,  Chenggong, Kunming 650500, \\ People's Republic of China \\
 $^{3}$ Department of Physics and Institute of Astronomy, National Tsing Hua University, 30013 Hsinchu,
 Taiwan \\
 }

\usepackage{amsmath}

\begin{document}

\date{Accepted 2023 April 3. Received 2023 February 28; in original form 2023 January 18 }

\pagerange{\pageref{0}--\pageref{0}} \pubyear{2022}

\maketitle

\label{firstpage}

\begin{abstract}
 We examine the time delay between radio and X-ray and between narrow radio frequency flares in Sagittarius A* (Sgr A*), from analyses of the synchrotron, bremsstrahlung 
  and monochromatic luminosity curves. Using the results of 2D relativistic radiation magnetohydrodynamic (MHD) simulations based on the shock oscillation model, we find three types of time delay between the synchrotron and bremsstrahlung emissions: Type A with a time delay of  2 -- 3 h on the shock
 descending branch, Type B with no time delay and Type C with an inverse time delay of  
 0.5 -- 1 h on the shock ascending branch.
The time delays in Types A and C are interpreted as a transit time of  Alfv\'{e}n and acoustic waves between  both emission dominant regions, respectively. 
The delay times between 22 and 43 GHz flares and between 8 and 10 GHz flares are  $\sim$ 13 -- 26 min and 13 min, respectively, while the inverse delay also occurs dependently on the shock location branch. 
These time delays between the narrow radio bands are interpreted as the transit time of the
 acoustic wave between the frequency-dependent  effective radii $R_{\tau_{\rm \nu=1}}$, at which the optical depth $\tau_{\rm \nu}$ at the accretion disc surface becomes $\sim$ unity. The shock oscillation model explains well the observed delay times of 0.5 -- 5 h between radio and X-ray, 20 -- 30 min between 22 and 43 GHz and $\sim$ 18 min between 8 and 10 GHz in Sgr A*.

\end{abstract}


\section{Introduction}
Sagittarius A* (Sgr A*),  the supermassive black hole at the center of our galaxy with a mass of $\sim$ 4 $\times$ 10$^6 M_{\odot}$ and located at 8.27 kpc away, have revealed very peculiar observations of
 too low luminosity and its complicated spectra  since its discovery (Genzel et al. 2003, Genzel, Eisenhauer \& Gillessen 2010). The observed luminosity is five orders of magnitude lower than that predicted by the standard thin disc model (Shakura \& Sunyaev 1973, hereafter, SS73 model)
 and the spectrum of Sgr A* differs from the multi-temperature black body spectra obtained from the  SS73 model.
  Various theoretical models which can explain too low luminosity and characteristic spectral properties have been proposed during the past three decades. Among them, advection dominated accretion flow (ADAF) model with high angular
 momentum is successful in explaining the observations well (see Narayan \& McClintock 2008; Yuan 2011; Yuan \& Narayan 2014, for review).

The latest observations in the range of radio to X-ray and  simultaneous observation campaigns through such energy bands
 showed  flares with a time scale of  $\sim$ hours to days  of Sgr A*  (Degenaar et al. 2013; Neilson et al. 2013, 2015; Ponti et al. 2015) and also time delays  between radio, IR, and X-ray emissions (Yusef-Zadeh et al. 2006a, 2008, 2009; Rauch et al. 2016; Ponti et al. 2017; Capellupo et al. 2017).
  Since the pioneering works of magnetized discs with shear instability (Balbus \& Hawley 1991; Hawley \& Balbus 1991), multidimensional magnetohydrodynamic (MHD) simulation works have shown that the magnetic field  and resultant magnetorotational instability (MRI) play important roles not only in the structure of accretion flow 
 but also the time-variations of luminosity and spectra around black holes (Machida, Hayashi \& Matsumoto 2000; Machida, Matsumoto \& Mineshige 2001; Stone \& Pringle 2001; Igumenshchev, Narayan \& Abramowicz 2003; Narayan, Igmenshchev \& Abramowicz 2003; Narayan et al. 2012; Yuan, Bu \& Wu 2012; Yuan et al. 2015). 
 In this respect, several MHD simulation works have attempted to address the rapid flares of Sgr A* (
Proga \& Begelman 2003; Chan et al. 2009; Dexter, Agol \& Fragile 2009; Yuan et al. 2009; Dodds-Eden et al. 2010; Ball et al. 2016; Ressler et al. 2017; Roberts et al. 2017; Li, Yuan \& Wang 2017).

 Besides the high angular momentum flow like ADAF, the low angular momentum flow models which exhibit the
 formation of standing shock near the event horizon have been
 developed to explain the quasi-periodic oscillations (QPOs) around black holes.
The studies of the standing shock in an astrophysical context were pioneered  by Fukue (1987) and then Chakrabarti (1989) and  further studies of the low angular momentum flows are refered to Okuda et al. (2019) and
Singh, Okuda \& Aktar (2021).
 Similarly to other MHD simulation works but based on the low angular momentum flow model, we examined the shock oscillation model for Sgr A* using 2D MHD and 2D radiation MHD simulations and showed that the magnetized flows yield large modulations of luminosities with a time-scale of $\sim$ 5 and 10 d (Okuda et al. 2019; Singh, Okuda \& Aktar 2021) and time delay of 1 -- 2 h between radio and X-ray flares of Sgr A* (Okuda, Singh \& Aktar 2022).
This paper is complementary to the time lag relation in the previous paper (Okuda, Singh \& Aktar 2022),
 where a simple two-temperature model with constant ratio of electron to ion temperatures is used.
 Here, we examine further time lag relation of flares  between radio and X-ray  and between narrow 
 radio frequency bands, using an improved two-temperature model where ion and electron temperatures 
 are obtained  by solving  the radiation energy equilibrium equation.

\section {Numerical method}
\subsection{Two-temperature model}
We use the previous numerical results of 2D relativistic radiation MHD simulations 
 for the long-term flares of Sgr A* (Okuda, Singh \& Aktar 2022) which were calculated using the RadRMHD module of the public library software ${\sc PLUTO}$ (Melon Fuksman \& Mignone 2019).
Accordingly, we know already the time variations of primitive variables of density $\rho$, velocity $\textbf{v}$,
 magnetic field  $\textbf {B}$, gas pressure $p_{\rm g}$, radiation energy density $E_{r}$ and radiation flux  $\textbf F$ obtained from a one-temperature model.    The computational domain is $0 \leq R \le 200$ and $-200 \le z \le 200$ with the resolution of $410 \times 820$ cells in the cylindrical coordinates (R, $\phi$, z), where $R$ and $z$ are expressed in the unit of the Schwarzschild radius $R_{\rm g}$ hereafter. 
The adiabatic index for studying the flow has been set as 1.6 for all simulation runs.

In this paper, we use a two-temperature model to evaluate the synchrotron and bremsstrahlung emissions,  solving the radiation energy equilibrium equation that Coulomb collisions transfer energy  rate  $q^{\rm ie}$ from ions to electrons equals to the sum of synchrotron cooling rate  $q_{\rm syn}$ and bremsstrahlung cooling rate $q_{\rm br}$ in electrons.
$q^{\rm ie}$ is given as follows (Stepney \& Guilbert 1983).

  \begin{eqnarray}
 &q^{\rm ie}&=5.61\times 10^{-32} {{n_{\rm e}n_{\rm i}(T_{\rm i}-T_{\rm e})}\over {K_2(1/\theta_{\rm e})
  K_2(1/\theta_{\rm i})}} \nonumber \\
  && \times \left[{{2(\theta_{\rm e}+\theta_{\rm i})^2+1}\over (\theta_{\rm e}+\theta_{\rm i})}
  K_1({{\theta_{\rm e}+\theta_{\rm i}}\over {\theta_{\rm e}\theta_{\rm i}}}) 
  +2K_0({{\theta_{\rm e}+\theta_{\rm i}}\over {\theta_{\rm e}\theta_{\rm i}}})\right] \nonumber \\
   &&  \hspace{3.8cm} {\rm erg \ cm^{-3} \; s^{-1}},
 \end{eqnarray}
  where $n_{\rm e}$ and $n_{\rm i}$ are the number density of electrons and ions, $K_0$, $K_1$
  and $K_2$ are
 modified Bessel functions, and the dimensionless electron and ion temperature are defined by

  \begin{equation}
  \theta_{\rm e}= {kT_{\rm e}\over {m_{\rm e}c^2}}, \;\;\; \theta_{\rm i}={kT_{\rm i}\over {m_{\rm p}c^2}},
  \end{equation}
where $k$, $m_{\rm e}$, $m_{\rm p}$ and $c$ are Boltzmann constant, electron mass,
 proton mass and the light velocity.

 The synchrotron cooling rate  $q_{\rm syn}$ is given by (Narayan \& Yi 1995; Esin et al. 1996)
\begin{eqnarray}
  q_{\rm syn}  &=& 
    {{2{\rm \pi} kT_{\rm e}\nu_{\rm c}^3 }\over {3Hc^2}}  
     +6.76\times 10^{-28} {n_{\rm i}\over{K_2(1/\theta_{\rm e})a_1^{1/6}}} \nonumber \\ 
  & \times &  [{1\over a_4^{11/2}}\Gamma({11\over 2},a_4\nu_{\rm c}^{1/3})
     +{a_2\over a_4^{19/4}} \Gamma({19\over 4},a_4\nu_{\rm c}^{1/3}) \nonumber \\        
  &+& {a_3\over a_4^4}( a_4^3\nu_{\rm c}+3a_4^2\nu_{\rm c}^{2/3}    
   +6a_4\nu_{\rm c}^{1/3}+6) {\rm e}^{-a_4\nu_{\rm c}^{1/3}}] \nonumber \\   
  && \hspace{3.8cm} {\rm erg \ cm^{-3} \ s^{-1}},
   \end{eqnarray}
   where $H$ is a scale height of the disc,
    \begin{eqnarray}
   a_1={2\over {3\nu_0\theta_{\rm e}^2}},\; a_2={0.4\over a_1^{1/4}},\; a_3={0.5316\over a_1^{1/2}},
   \; a_4={1.8899a_1^{1/3}}, \nonumber \\                                    
   \Gamma(a,x)= \int_{x}^{\infty}t^{a-1}{\rm e}^{-t}dt, 
   \nu_0={eB\over {2{\rm \pi} m_{\rm e}c}} {\; \rm and} \;\nu_{\rm c}={3\over 2}\nu_0\theta_{\rm e}^2x_{\rm M}.
     \end{eqnarray}
     Here, $e$ and $B$ are the electron charge and the strength of the magnetic field and $x_{\rm M}$  is determined from the next equation
     as
    \begin{eqnarray}
      {\rm exp}(1.8899x_{\rm M}^{1/3} ) &=&  2.49\times 10^{-10}{4{\rm \pi} n_{\rm e}R\over
       B} {1\over {\theta_{\rm e}^3 K_2(1/\theta_e) }} \nonumber \\
       &\times&    \left({1\over  x_{\rm M}^{7/6} }
        + {0.40 \over x_{\rm M}^{17/12}}+{0.5316\over {x_{\rm M}^{5/3}}}\right). \nonumber \\        
    \end{eqnarray}

 The bremsstrahlung cooling rate $q_{\rm br}$ is given as follows (Stepney \& Guilbert 1983).
   \begin{eqnarray}
  q_{\rm br} = q_{\rm ei} + q_{\rm ee},
  \end{eqnarray}
  
  \begin{eqnarray}
  q_{\rm ei}=1.48\times 10^{-22}n_{\rm e}^2 F_{\rm ei}(\theta_{\rm e}) \hspace{1cm}{\rm erg \ cm^{-3}\; s^{-1}},
  \end{eqnarray}
  
  \begin{eqnarray}
   \;\;\;F_{\rm ei}(\theta_{\rm e}) =\left\{\begin{array}{ll}
    1.02\theta_{\rm e}^{1/2}(1+1.78\theta_{\rm e}^{1.34}) & \mbox{ for $\theta_{\rm e} <1$},  \\
    1.43\theta_{\rm e}[{\rm ln}(1.12\theta_{\rm e}+0.48)+1.5]
                                                          & \mbox{ for  $\theta_{\rm e}>1$},
                                              \end{array}
                                      \right. 
  \end{eqnarray}
  
  \begin{eqnarray}
  q_{\rm ee} = \left\{\begin{array}{llll}
     2.56\times 10^{-22}n_{\rm e}^2\theta_{\rm e}^{1.5}
       (1+1.10\theta_{\rm e}+\theta_{\rm e}^2-1.25\theta_{\rm e}^{2.5})  & \\
     \hspace{3cm}  {\rm erg \ cm^{-3} \; s^{-1} }  \mbox{ for $\theta_{\rm e} <1$}, 
  & \\
    3.40\times 10^{-22}n_{\rm e}^2\theta_{\rm e} [{\rm ln}(1.123\theta_{\rm e})+1.28 ]
  & \\                                                                                       
     \hspace{3cm}   {\rm erg \ cm^{-3} \; s^{-1}}  \mbox{ for  $\theta_{\rm e} >1$}. & \\
                               \end{array}
                      \right.
    \end{eqnarray}

 Furthermore, from matching of one-temperature model to two-temperature model, we have, 
 \begin{equation} 
   T_{\rm e} + T_{\rm i} = 2T,
 \end{equation}
 because we set the gas pressure $p_{\rm g}$ = $2nkT$ = $n_{\rm e}kT_{\rm e} + n_{\rm i}kT_{\rm i}$ and $n = n_{\rm e} = n_{\rm i}$, where $T$ and $n$
 are the temperature and the number density, respectively,  in the one-temperature model.
We solve numerically the equations of $q^{\rm ie} =  q_{\rm syn} + q_{\rm br}$ and $T_{\rm e} + T_{\rm i} = 2T$
 and obtain ion temperature $T_{\rm i}$ and electron temperature $T_{\rm e}$.

\begin{figure}
\begin{center}
        \includegraphics[width=0.4\textwidth]{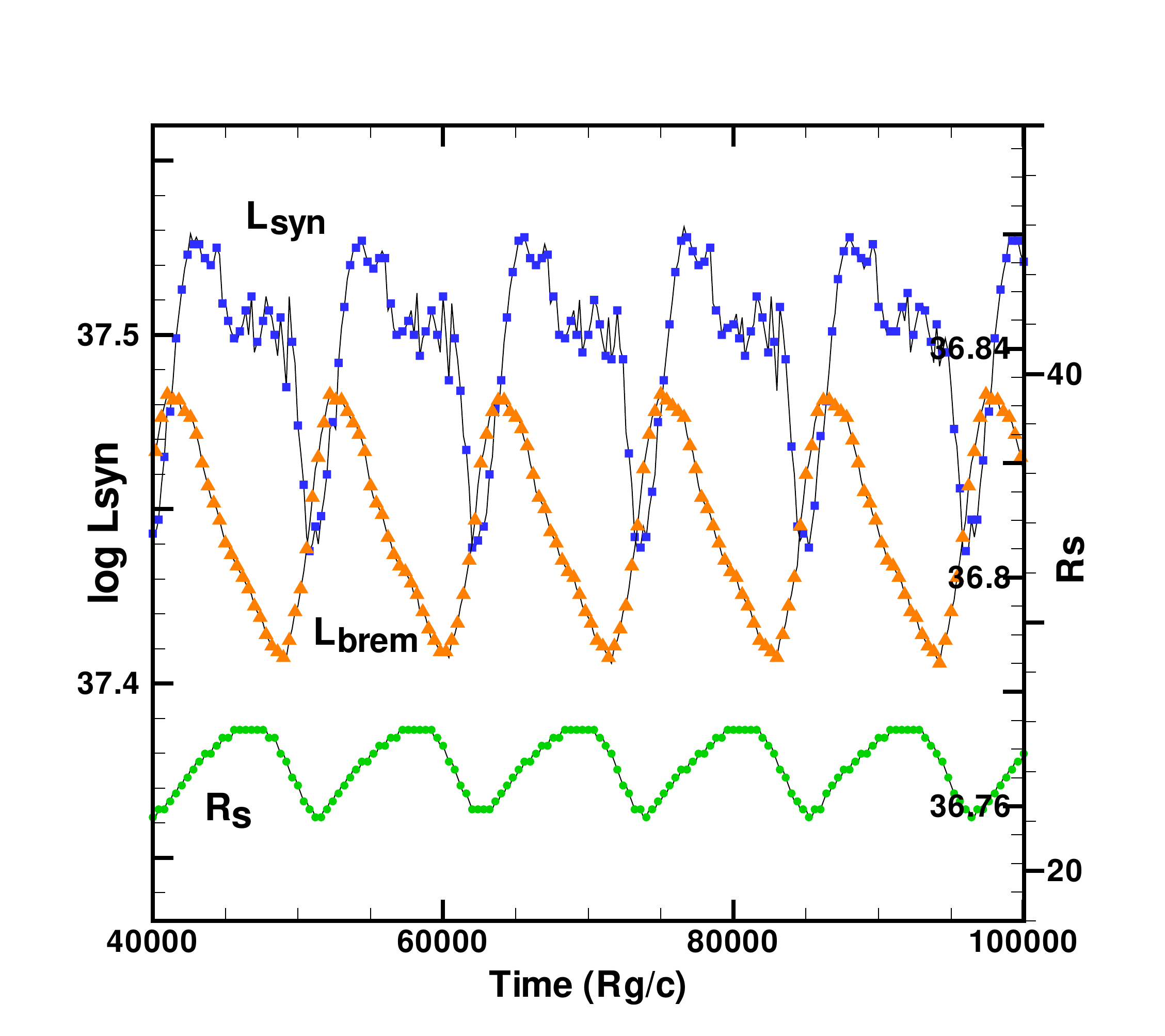}
\caption {Time variations of synchrotron luminosity $L_{\rm syn}$ (blue square), bremsstrahlung luminosity $L_{\rm brem}$ (orange Delta)  (erg s$^{-1}$) and oscillating shock location $R_{\rm s}$ (green circle) on the equator during $t= 4\times 10^4 - 1.0\times 10^5$ in unit of $R_{\rm g}/c$ for model A with $\lambda$= 1.65,  $\epsilon =6.89 \times 10^{-3}$ at the outer boundary $R_{\rm out}$ = 50 and the magnetic field strength $\beta_{\rm out}$ = $10^4$. 
The scale of $L_{\rm brem}$ is shown inside on the right vertical axis
The circle points on the curves  are plotted every time interval of $200 R_{\rm g}/c$ ($\sim 0.02$) s.
  }
\end{center}
\end{figure}

\begin{figure}
\begin{center}
\includegraphics[width=0.4\textwidth]{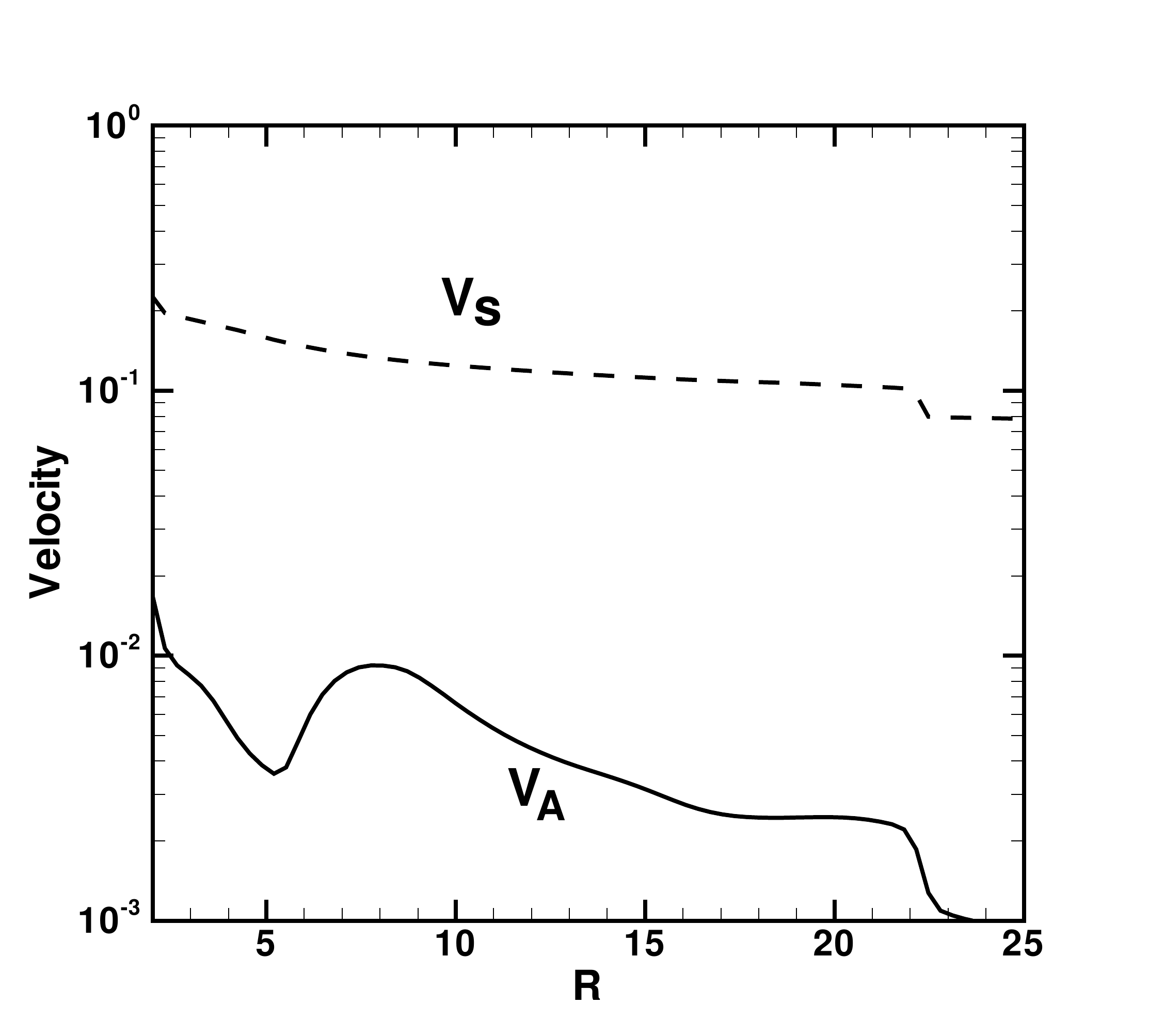}
\caption {Alfv\'{e}n velocity $V_{\rm A}$ (solid line) and sound velocity $V_{\rm s}$ (dashed line)  on the equator at $t= 4\times 10^4 (R_{\rm g}/c)$ for model A.
 Here, the sound velocity is $\sim 0.1c$ in the region of $R \leq 20 $, while the Alfv\'{e}n velocity is roughly
 $\ge$ 0.003$c$.
 }
\end{center}
\end{figure}

 We confirm here the optical thickness  $\Delta\tau =\kappa \rho \Delta R$ across the mesh size $\Delta R$
  for the present model for Sgr A* by (see Okuda, Singh \& Aktar, 2022),
\begin{eqnarray}
  \Delta\tau  \sim 2\times 10^{-28}\left(\frac{\rho}{10^{-16}}\right)^2 \left(\frac{T}{10^9}\right)^{-3.5} \left(\frac{\Delta R}{0.2}\right) \ll 1,
 \end{eqnarray}
 where the opacity $\kappa$ is given by the Kramers approximation corresponding to the bremsstrahlung emission.
 Accordingly, the gas is fully optically thin to the bremsstrahlung emission but not to the synchrotron emission throughout the region considered here. 
 If we consider the monochromatic radiation in the radio frequencies where the synchrotron
  radiation is dominant, the gas may be optically thick or thin because the opacity becomes large dependently 
 on $\nu$ and the location $R$. 
  In PLUTO code, we use the one-temperature model and only the radiative energy loss by the fre-free
 emissions. 
  Then, the total radiative luminosity $L_{\rm rad}$ corresponds to the free-free emissions and  is evaluated at the outer z-boundary and the outer R-boundary surfaces, as follows
 \begin{eqnarray}
   L_{\rm rad} =\int {\bf F} {\rm d} {\bf S},
 \end{eqnarray}
where ${\bf F}$ is the radiation energy flux at the boundary surfaces.
On the other hand, the total luminosities $L_{\rm syn}$ and $L_{\rm brem}$ due to the synchrotron and bremsstrahlung emissions, respectively, are approximately given

\begin{eqnarray}
 L_{\rm brem} =\int q_{\rm brem} {\rm d} V,
\end{eqnarray}

\begin{eqnarray}
  L_{\rm syn} =\int q_{\rm syn} {\rm d} V, 
\end{eqnarray}
where the volume integration is carried out over all computational zones.
The expression for $L_{\rm syn}$ may be not exact because the synchrotron emission becomes
 optically thick at a limited range of the frequencies in the inner region of the disc as  shown in later numerical results (Figs.~10 -- 11) but gives important amounts of radiation energy in the radio bands
 from the viewpoint of the time-delay analysis between the radio and X-ray emission peaks.

\subsection{Monochromatic radiation energy flux $F_{\rm \nu}$ and luminosity $L_{\rm \nu}$}
To examine time delay between flares at narrow radio frequency band such as 22, 43 and 350 GHz, we calculate monochromatic radiation flux $F_{\rm \nu}$ at the disc surface and luminosity $L_{\rm \nu}$. Assuming a locally plane-parallel approximation for the accretion disc flow, the radiation flux $F_{\rm \nu}$ at a given radius R is given by  (see Manmoto, Mineshige \& Kusunose 1997)
\begin{eqnarray}
 F_{\rm \nu }(R) = {{2\pi}\over \sqrt{3}} B_{\rm \nu} \left[1 - {\rm exp} (-2 \sqrt{3} {\tau_{\rm \nu}}^*)\right] 
 \;\; {\rm erg}\;{\rm cm}^{-2}\; {\rm s}^{-1}{\rm Hz}^{-1},
\end{eqnarray}
where $B_{\rm \nu}$ is the Planck function,  ${\tau_{\rm \nu}}^* \equiv (\pi^{1/2}/2) \kappa_{\rm \nu}(0) H$ is the optical depth at the accretion disc surface and $\kappa_{\rm \nu}(0)$ is the absorption coefficient on the equatorial plane. Assuming LTE, $\kappa_{\rm \nu} = \chi_{\rm \nu}/(4 \pi B_{\rm \nu})$, where $\chi_{\rm \nu}$ is the emissivity. Since we are interested in the emissions at radio bands such as 22, 43 and 350 GHz, 
 the relation (15) is valid for the gas which includes optically thick disc region.
 We consider the synchrotron emissivity  $\chi_{\rm \nu,syn}$ by a relativistic Maxwellian distribution of electrons  as follows (Narayan \& Yi 1995)

\begin{eqnarray}
 \chi_{\nu,syn} = 4.43 \times 10^{-30} { {4\pi n_{\rm e} \nu} \over {K_{2}(1/\theta_{\rm e})}} I(x) 
 \;\; {\rm erg}\;{\rm cm}^{-3}\;{\rm s}^{-1} {\rm Hz}^{-1},
\end{eqnarray}
where $x \equiv {4\pi m_{\rm e}c\nu}/({3eB{\theta_{\rm e}}^2})$,
 \begin{eqnarray}
      I(x) ={4.0505\over x^{1/6}} \left(1+ {0.40 \over x^{1/4}}+{0.5316\over x^{1/2}}\right) {\rm exp} (-1.8899x^{1/3}).     
 \end{eqnarray}
The monochromatic luminosity $L_{\rm \nu}$ is given by 

\begin{eqnarray}
  L_{\rm \nu} =\int 2\pi R F_{\rm \nu}(R) dR,
\end{eqnarray}
where the integration is carried out up to the outer radial boundary.

\section{Numerical results}

\subsection{Application to a typical oscillating shock model under weak magnetic field}
 In the previous paper (Okuda et al. 2019), to find general effects of the magnetic field on 2D hydrodynamical flow with standing shock around a black hole of mass 10 $M_{\odot}$, we examined
  a  2D magnetohydrodynamical flow (hereafter model A) with a typical set of  flow parameters of  the specific angular momentum $\lambda$= 1.65, the specific energy $\epsilon =6.89 \times 10^{-3}$ at the outer boundary $R_{\rm out}$ = 50. 
 The result shows  steady standing shocks under a very weak magnetic field with $\beta_{\rm out}$=
 $10^5$ and $10^9$, a periodic shock oscillation between  $R$ = 22 -- 25 under a weak magnetic field with $\beta_{\rm out}=10^4$, and a chaotic variation of the shock under a strong magnetic field with 
$\beta_{\rm out}= 4\times 10^3$, where $\beta_{\rm out}$ is the ratio of gas pressure to
magnetic pressure at the outer R-boundary $R_{\rm out}$. Applying the present two-temperature model  to the above second case with $\beta=10^4$, we examined how the time delay between the synchrotron and bremsstrahlung emissions occurs, because this is a simple example to understand the time lag relation.
 Fig.~1 shows the time variations of synchrotron luminosity $L_{\rm syn}$ (blue square),  bremsstrahlung luminosity $L_{\rm brem}$ (orange Delta) (erg s$^{-1}$) and  standing shock location $R_{\rm s}$ 
(green circle) on the equator for  model A, where the curves are plotted every time interval of 200 $R_{\rm g}/c$ ($\sim 0.02$ s). Hereafter, we term the plotted time interval as one point
 time interval.
 The bremsstrahlung luminosity begins to increase after the maximum shock location and becomes maximum 
after the shock contracts minimally because the temperature, density and magnetic field strength behind the shock are mostly enhanced in the innermost region. In Fig.~1, the synchrotron luminosity peak delays the bremsstrahlung one by $\sim $ 8 --10 points time interval. 
 From examining the radial dependence of the synchrotron and bremsstrahlung volume emissions, 
 we find that most of the synchrotron luminosity  is emitted in the inner region of $R \le R_{\rm syn}$ (= 5) at all phases.
 While the bremsstrahlung emission dominant region is confined at $R \sim R_{\rm brem}$ 
(=10).
Therefore,  even after the bremsstrahlung luminosity is attained to the maximum, the perturbations of acoustic and Alfv\'{e}n waves at  $R \sim R_{\rm brem}$ influence the downstream flow. 
 As a result, the synchrotron luminosity peak delays the bremsstrahlung one by a 
 transit time of Alfv\'{e}n wave between $R_{\rm syn}$  and $R_{\rm brem}$, because 
 the synchrotron emission depends strongly on the magnetic field.

 Fig.~2 shows Alfv\'{e}n velocity $V_{\rm A}$ (solid line) and sound velocity $V_{\rm s}$ (dashed line) on the 
equator at $t= 4\times 10^4 R_{\rm g}/c$ near the bremsstrahlung luminosity peak in model A, 
where  $V_{\rm s} \sim 0.1c$  in the inner region of $R \le 20$, while $V_{\rm A} \ge$ 0.003$c$. 
 The delay time of the synchrotron luminosity peak to the bremsstrahlung one is  estimated as $ (R_{\rm brem} -R_{\rm syn}) / V_{\rm A} =  1.67 \times 10^3 R_{\rm g}/c$ ($\sim$ 8 points time interval), which agrees well  with the above time delay of 8 -- 10 points time 
interval between the synchrotron and bremsstrahlung luminosity peaks.
  Also, the synchrotron luminosity dip delays the bremsstrahlung one
 nearly by $\sim$ 8 points time interval, because $R_{\rm brem}$ and $R_{\rm syn}$ are not so variable.
It should be noted that, in the accretion flow, the gas always flows inwards  near the equator and the outflow appears only along the rotational axis.
This shows that the time delay occurs without any expanding hot blob which is discussed
 later with relevance to the plasmon model.

\begin{figure*}
    \begin{tabular}{cc}

      \begin{minipage}{0.4\linewidth}
        \centering
        \includegraphics[width=0.9\textwidth]{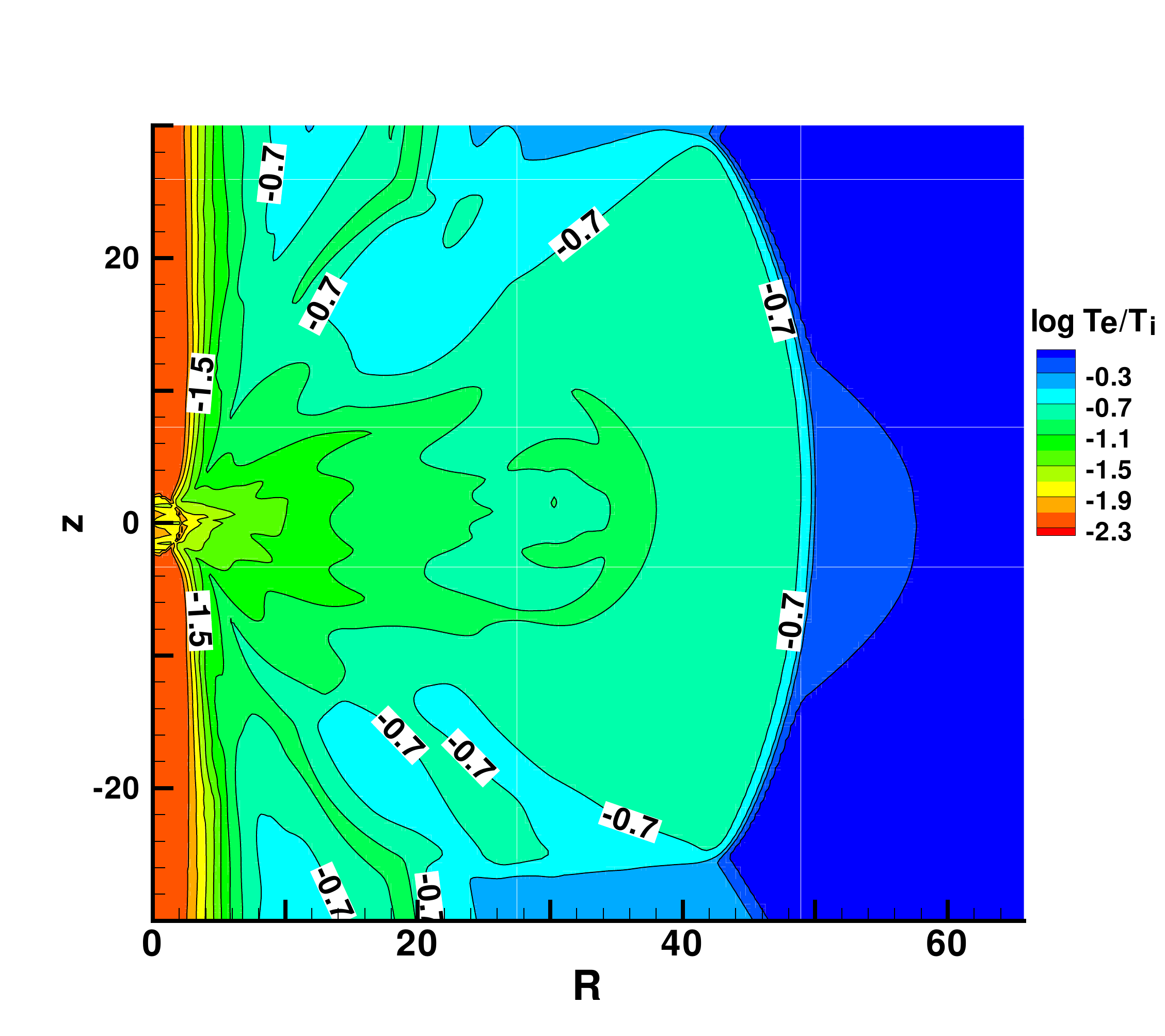}
        \label{fig:3a}
      \end{minipage}

      \begin{minipage}{0.4\linewidth}
        \centering
        \includegraphics[width=0.9\textwidth]{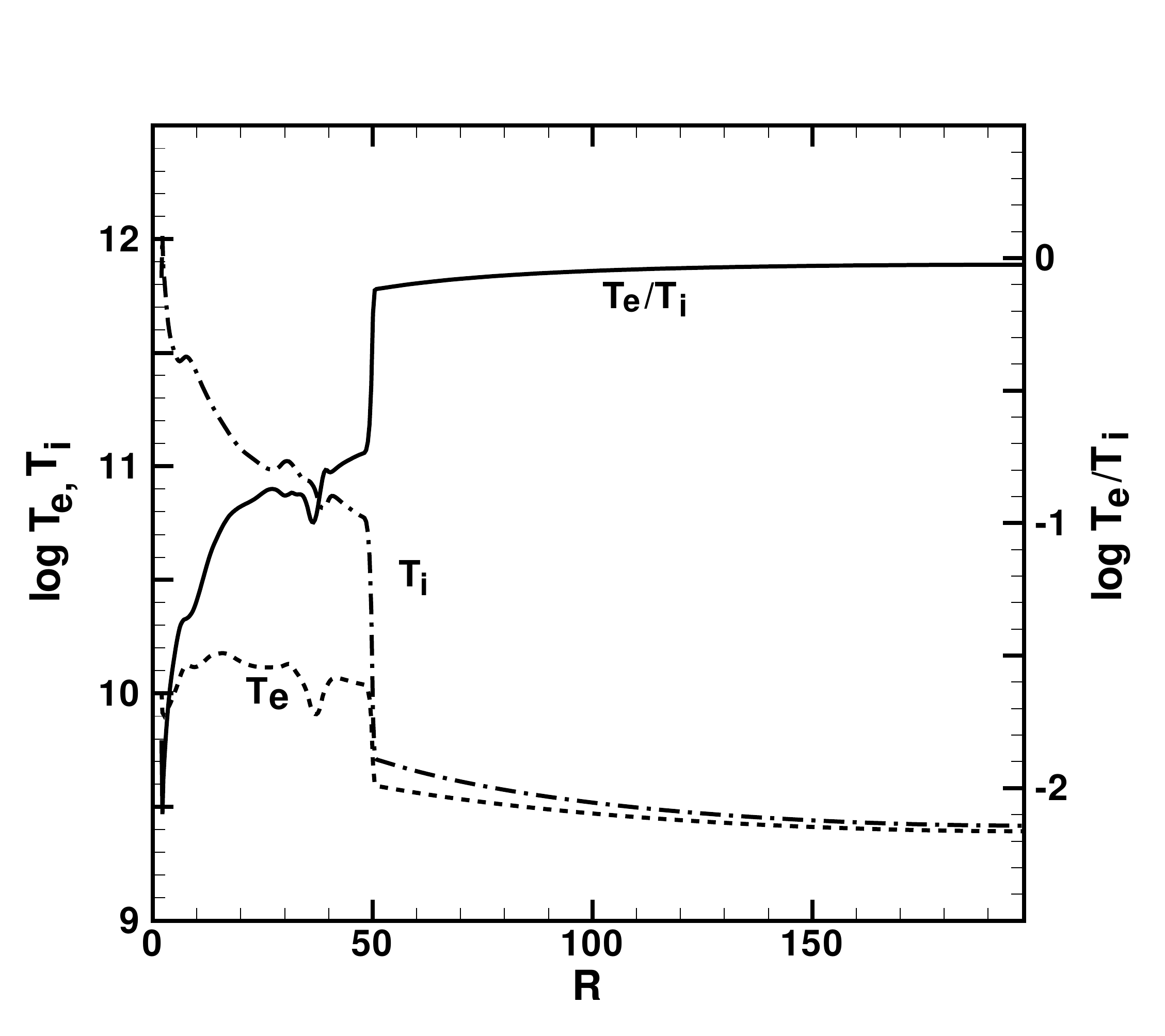}
        \label{fig:3b}
      \end{minipage} \\

    \end{tabular}
\caption {(Left-hand panel): 2D contours of ratio $T_{\rm e}/T_{\rm i}$ of electron temperature $T_{\rm e}$ to ion temperature $T_{\rm i}$ at $t= 1.98\times 10^7$ s for model Rad1. (Right-hand panel):  1D profiles of $T_{\rm e}$ (small dash line), $T_{\rm i}$ (dash-dot line) and  ratio $T_{\rm e}/T_{\rm i}$ (solid line)  on the equator. The shock location on the equator is found at $R\sim 50$.
  }
  \label{Fig: fig3}
  \end{figure*}

\begin{figure*}
    \begin{tabular}{cc}

      \begin{minipage}{0.8\linewidth}
        \centering
        \includegraphics[width=100mm,height=80mm]{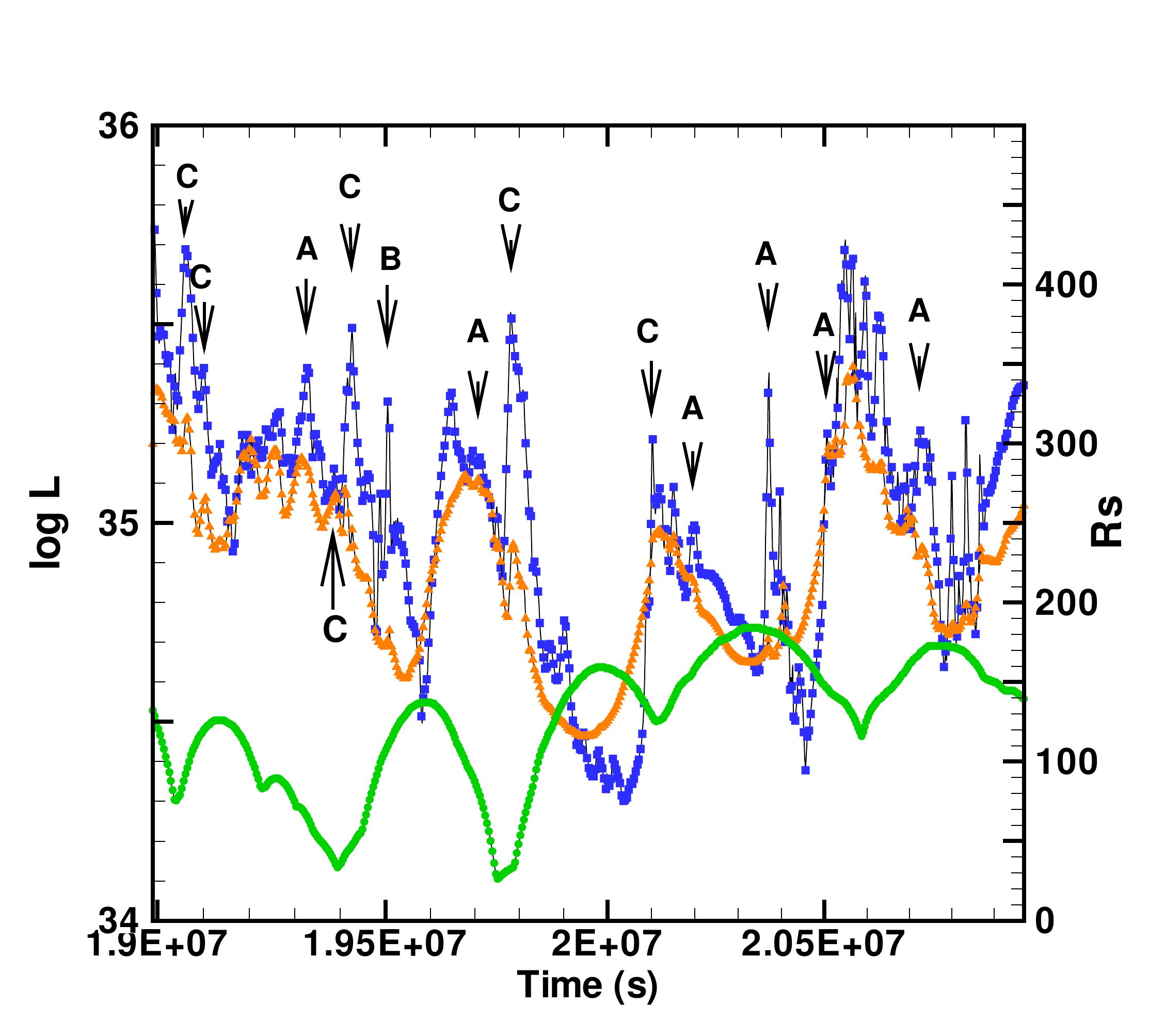}
        \label{fig:4}
      \end{minipage}

      \begin{minipage}{0.0\linewidth}
        \centering
        \includegraphics[width=0.8\textwidth]{fig4.pdf}
        \label{fig:4}
      \end{minipage} \\

    \end{tabular}
\caption{Time variations of synchrotron luminosity $L_{\rm syn}$ (blue square), bremsstrahlung luminosity $L_{\rm brem}$ (orange Delta)  (erg s$^{-1}$) and oscillating shock location $R_{\rm s}$  (green circle) on the equator during $t=  1.9 - 2.1 \times 10^7$ s in model Rad1. The circle points on the lines  are plotted every time interval of 50 $R_{\rm g}/c$ ($\sim$ 0.55 h). The arrow shows the local maximum points on the synchrotron luminosity curve and the marks of A, B and C show delay types: Type A with a positive time delay of  the synchrotron to bremsstrahlung flare, Type B with no time delay  and Type C with a negative time delay
 at which the bremsstrahlung flare delays the synchrotron one. 
  }
  \label{Fig: fig4}
  \end{figure*}

\subsection{Application to Sgr A* }
 In this section, we apply the present two-temperature model to the numerical results of 2D relativistic radiation MHD simulations for the long-term flares of Sgr A*.
 We consider  the mass of Sgr A* as $M = 4\times 10^6 M_{\odot}$.
 Here, the unit of distance $R_{\rm g}$ is $2GM/c^2 \sim 1.19\times 10^{12}$ cm.

\subsubsection{Time delay between radio and X-ray flares}
To examine the time delay between the synchrotron and bremsstrahlung emissions from the radiation MHD simulations, we select time evolutions during 1.90 - 2.1 $\times 10^7$ s  at model Rad1 in the previous paper (Okuda, Singh \& Aktar 2022) and recalculated the simulations using a smaller time step of 50$R_{\rm g}/c$ 
 ( $\sim$ 0.55 h) in order to get better time resolution for the time delay analysis.
 First, we show  the electron and ion temperatures obtained from the two-temperature model and compare them with other results in the advection-dominated two-temperature model. Fig.~3 shows  2D contours of $T_{\rm e}/T_{\rm i}$ (left-hand panel) and 1D profiles of $T_{\rm e}$ (small dash line), $T_{\rm i}$ (dash-dot line) and $T_{\rm e}/T_{\rm i}$ (solid line)  on the equator (right-hand panel) at $t = 1.98 \times 10^7$ s for 
 Rad1 case. Here, the shock location on the equator is found at $R \sim 50$.
The left panel  shows that the electron temperatures are not so different from the ion temperatures in
 far distant region from the center but the ratio $T_{\rm e}/T_{\rm i}$ becomes small as 0.01 -- 0.1 in the inner region of $R \le$ 20 and along the rotational axis.
 In the right panel, the temperature distributions of $T_{\rm e}$ and $T_{\rm i}$ are qualitatively similar to those in the steady solutions of advection-dominated accretion flow without shock by Manmoto, Mineshige \& Kusunose (1997) and also
of advective flows with shocks by  Mandal \& Chakrabarti (2005) and Dihingia, Das \& Mandal (2018), except that
the difference between $T_{\rm e}$ and $T_{\rm i}$ becomes very large across the shock in our case.
Then, we may regard the two-temperature model to be reasonable.

The radiative luminosity $L_{\rm rad}$ is evaluated at the vertical outer z-boundaries and the  outer R-boundaries, while the synchrotron and bremsstrahlung emissions are effectively emitted near the equatorial plane.
Accordingly, the former ($L_{\rm rad}$) is expected to delay the latters ($L_{\rm syn}$ and $L_{\rm brem}$) by a transit time of the light from the equator to the outer z-boundary 
if there is no time lag between the synchrotron and bremsstrahlung emissions on the equator. 
The transit time is 200$R_{\rm g}/c$ ($7.9\times 10^3$ s $\sim$ 2 h).
 Actually, we confirm that $L_{\rm rad}$ quite often delays $L_{\rm brem}$ by $\sim$ 2 h at local maximum and minimum points of the luminosity curves
 but that the delay is different for $L_{\rm syn}$ because the synchrotron emission delays
 the bremsstrahlung emission in most flares.
 In our radiation MHD simulations using the RadRMHD module of the ${\sc PLUTO}$ code, it should be noted that  the radiative cooling is taken into account only the free-free emission between the ions and electrons because the one-temperature model is used in the ${\sc PLUTO}$ code. Then the radiative luminosity is underestimated compared with the bremsstrahlung luminosity due to the two-temperature model,
 since in the latter the electron-electron bremsstrahlung is also considered in addition to the electron-proton bremsstrahlung. However, it will not affect our time delay analysis in this paper.

 Fig.~4 shows the time variations of the synchrotron luminosity $L_{\rm syn}$ (blue square),
 bremsstrahlung luminosity  $L_{\rm brem}$ (orange Delta) and  oscillating shock location $R_{\rm s}$ (green circle on the equator during $t=  1.9 - 2.1 \times 10^7$ s in model Rad1.
 The curves are plotted every $50 R_{\rm g}/c$ time interval (1.98$\times 10^3$ s $\sim$ 0.55 h).
 The attached ``arrow" indicates local peaks on the synchrotron luminosity curve
 and the marks of A, B and C expresses delay types; Type A with a positive time delay
 of the synchrotron to bremsstrahlung flare, Type B with no time delay  and Type C with a negative time delay
 at which the bremsstrahlung flare delays the synchrotron one.
 From the figure, we can find roughly the distribution of  delay-type of the flares. 

\begin{figure*}
    \begin{tabular}{cc}

      \begin{minipage}{0.8\linewidth}
        \centering
        \includegraphics[width=0.8\textwidth]{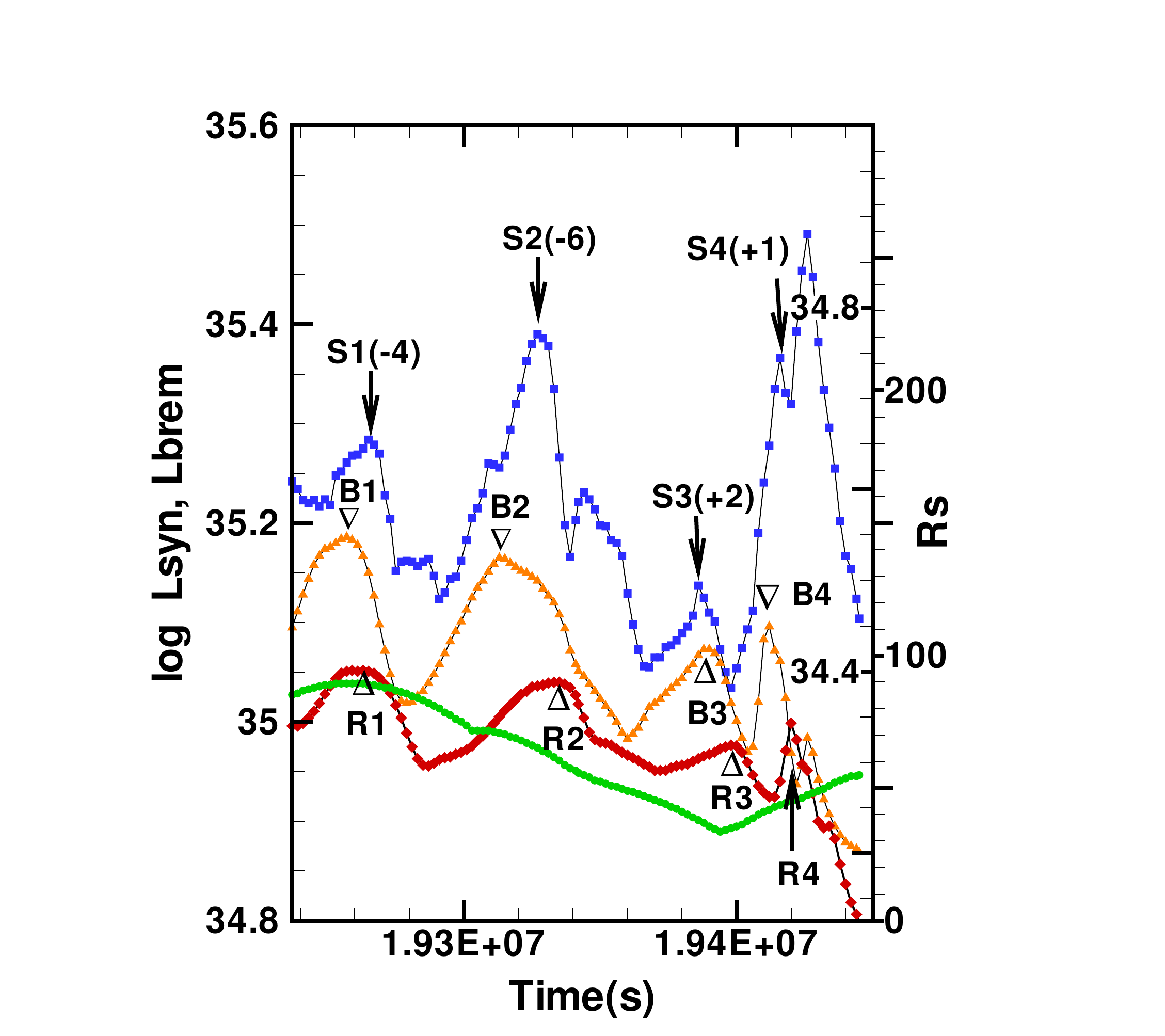}
        \label{fig:5}
      \end{minipage}

      \begin{minipage}{0.0\linewidth}
        \centering
        \includegraphics[width=0.8\textwidth]{fig5.pdf}
        \label{fig:5a}
      \end{minipage} \\

    \end{tabular}
\caption{Details of the luminosity curves of  $L_{\rm rad}$ (red diamond), $L_{\rm syn}$ (blue square),  $L_{\rm brem}$ (orange Delta) and oscillating shock location $R_{\rm s}$ (green circle) on the equator during $t$= 1.9237 -- 1.9445 $\times 10^7$ s in model Rad1. The symbols ``arrow", ``$\bigtriangledown$" and ``$\bigtriangleup$"
 show the corresponding peaks in the curves of  $L_{\rm syn}$, $L_{\rm brem}$ and  $L_{\rm rad}$,
 respectively. The letters ${\textbf Si}$, ${\textbf Bi}$ and  ${\textbf Ri}$  express the ${\textbf i}$ th flare event on the  synchrotron, bremsstrahlung and radiative luminosity curves, respectively.
 The number within the bracket of the synchrotron symbol S shows the delay point number 
  relative to the bremsstrahlung emission. For example, ${\textbf S1(-4)}$ shows that the 1 th synchrotron  flare event delays the bremsstrahlung emission by 4 points time interval ($\sim$ 2 h). The scale of $L_{\rm rad}$ is shown inside on the right vertical axis.
  }
  \label{Fig: fig5b}
  \end{figure*}

\begin{figure*}
    \begin{tabular}{cc}

      \begin{minipage}{0.8\linewidth}
        \centering
        \includegraphics[width=0.8\textwidth]{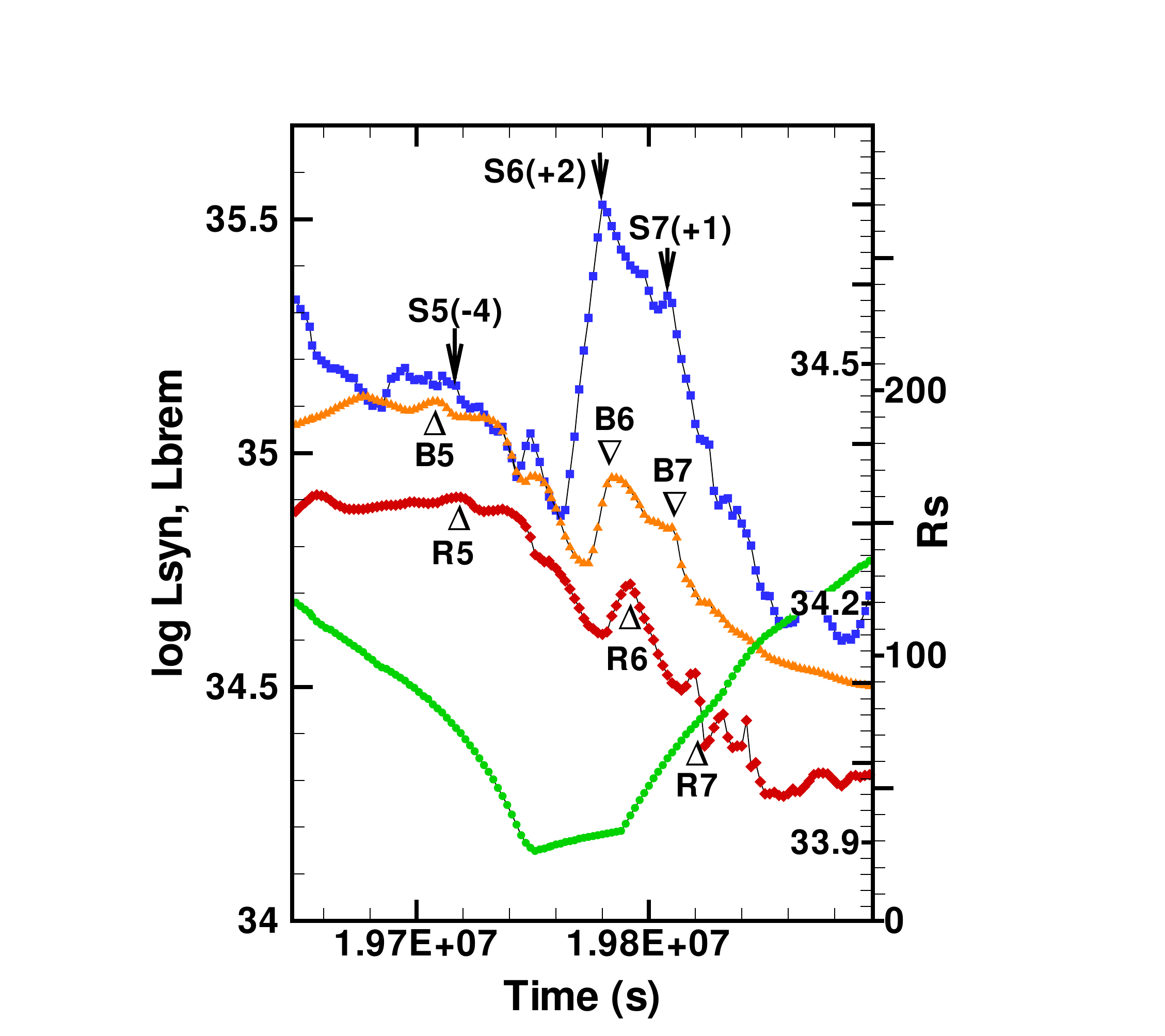}
        \label{fig:6}
      \end{minipage}

      \begin{minipage}{0.0\linewidth}
        \centering
        \includegraphics[width=0.8\textwidth]{fig6.pdf}
        \label{fig:6}
      \end{minipage} \\
    \end{tabular}
\caption{Same as Fig.~5 but during  $t$= 1.965 -- 1.990 $\times 10^7$ s.
  }
  \label{Fig: fig6}
  \end{figure*}

 In order to show  details of the time delay, 
  we plot typical two cases of the luminosity curves $L_{\rm rad}$ (red diamond), $L_{\rm syn}$ (blue square) and $L_{\rm brem}$
(orange Delta) over a  shock oscillation cycle in Figs.~5 and~6.
The luminosity curve  $L_{\rm rad}$ is useful to distinguish
 the peak and dip  points in the bremsstrahlung luminosity curve because $L_{\rm rad}$ delays 
 nearly always $L_{\rm brem}$  by 200$R_{\rm g}/c$, that is, 4 points time interval.
 To be able to distinguish the time delay between $L_{\rm rad}$, $L_{\rm syn}$ and $L_{\rm brem}$ clearly, the scale of $L_{\rm rad}$ is shown inside on the right vertical axis.
 In the figures, the symbols ``arrow", ``$\bigtriangledown$" and ``$\bigtriangleup$"
 show the corresponding peaks in the curves of  $L_{\rm syn}$, $L_{\rm brem}$ and  $L_{\rm rad}$,
 respectively,  and the letters ${\bf Si}$, ${\bf Bi}$ and  ${\bf Ri}$  express the ${\bf i}$ th flare event in the  synchrotron, bremsstrahlung and radiative luminosity curves, respectively 
and the number within the bracket of the synchrotron symbol shows the delay point 
  number relative to the bremsstrahlung emission. For example, ${\bf S1(-4)}$ shows that the 1 th
 synchrotron  flare event delays the bremsstrahlung emission by 4 points time interval ($\sim$ 2 h).
 Type A flares (events 1, 2 and 5)  occur on the shock descending branch and have a delay of 4 -- 6 points time interval (2 -- 3 h).
 Type B is the intermediate case between Type A and Type C and  occurs only when the shock is far away from the center.
 Type C flares (events 3, 4, 6 and 7) with a delay time of  1 -- 2 points time interval (0.5 -- 1 h)  occurs when the
 oscillating shock attains the minimum location and expands.

In the same way as the previous paper (Okuda, Singh \& Aktar 2022), to examine the effective emitting regions of $L_{\rm syn}$ and $L_{\rm brem}$, we calculate local synchrotron and bremsstrahlung luminosities $L_{\rm syn}(R)$  and $L_{\rm brem}(R)$ which are emitted within a sphere of radius $R$. As a result,  we confirmed that most of the synchrotron emission is emitted within a confined small region of  $R \le$ 5 and 60 -- 80 
 percent of the bremsstrahlung emission comes from the inner region of $10 \le R \le 20$ 
 at phases of the minimum shock location.
 Then, we regard the synchrotron and bremsstrahlung emission dominant radii  $R_{\rm syn}$ and
 $R_{\rm brem}$ as 5 and 10, respectively.
 The delay time between the synchrotron and bremsstrahlung emissions in Type A is estimated
 as the transit time of the Alfv\'{e}n wave from $R_{\rm brem}$ to $R_{\rm syn}$.
 The acoustic wave also contributes to the synchrotron radiation  but
 is not effective as the Alfv\'{e}n wave because the synchrotron emission strongly depends on
 the  magnetic field. 
 Fig.~7 shows profiles of radial velocity $V_{\rm r}$ (solid line), sound velocity $V_{\rm s}$ (dashed line) and Alfv\'{e}n velocity $V_{\rm A}$ (dash-dot-dot line) on the equator at $t = 1.936 \times 10^7$ s (left-hand panel) and  $t = 1.980 \times 10^7$ (right-hand panel).
 The oscillating shock locations are found at $R \sim$ 48 on the shock descending branch (left-hand panel) and $\sim$ 50 on the shock ascending branch (right-hand panel). 
Here, $V_{\rm s} \ge 0.1c$ and $V_{\rm A}  \ge 0.02c$ at $R \le 20$ on the shock descending branch.
 Then, the transit time ($R_{\rm brem}-R_{\rm syn})/V_{\rm A}$ in Type A is estimated as 
 $\le 250 R_{\rm g}/c$ = 2.5 h which corresponds well to the above delay time 2 -- 3 h.

 Type B flare with no time-lag appearing at the far distant shock location is naturally understood from the above consideration. Near the maximum shock location, the synchrotron and bremsstrahlung
 dominant regions are as $R_{\rm syn} \sim 5$ and $R_{\rm brem} \ge 50$, respectively, the Alfv\'{e}n velocity is small as $\sim 10^{-3}$ in the distant region and the shock strength is weak. Then the transit time of the Alfv\'{e}n wave in Type B is too long compared with the above transit time in Type A. 
 Thus, in Type B,  there is no interaction of the perturbed Alfv\'{e}n wave between the synchrotron and bremsstrahlung dominant regions, that is, there is no time lag between the synchrotron and bremsstrahlung emissions on the equator.

 Type C flare appears in a special state as follows.
 During the time evolution of the magnetized flow under MRI, the magnetic field intermittently increases and decreases near
 the event horizon, and the magnetic pressure gradient force begins to dominate the gas pressure gradient force, the gravitational and centrifugal forces along the rotational axis and  sometimes 
in the equatorial direction. This leads to an intermittent high-velocity jet along the rotational axis and an outflow even in the equatorial direction. The outflow above and below the equator is at one time faded into the strong accreting flow, at other times leads to turbulent flow and  grows
 in the outer turbulent flow as an expanding hot and dense blob as is explained  in 
 Figs.~7, ~8 and ~12 in the previous paper (Okuda, Singh \& Aktar 2022).
The expanding hot blob sometimes interacts with the contracting oscillating shock and is incorporated into the shock, a strong flare occurs as Type C.
 The time delay of Type C is estimated as the transit time of the perturbed acoustic wave
 from $R_{\rm syn}$ to $R_{\rm brem}$ because the powerful  synchrotron emission influences 
 the outer bremsstrahlung dominant region through the acoustic wave, contrary to
  Type A, because the bremsstrahlung emission depends only on the density and the temperature. 
 The delay time is $\sim (R_{\rm brem}-R_{\rm syn})/V_{\rm s} \le 50 R_{\rm g}/c$ (0.55 h), 
 using $V_{\rm s} \ge 0.1c$, which is very small compared with that in Type A.

\begin{figure*}
    \begin{tabular}{cc}
      \begin{minipage}{0.45\linewidth}
        \centering
        \includegraphics[width=0.8\textwidth]{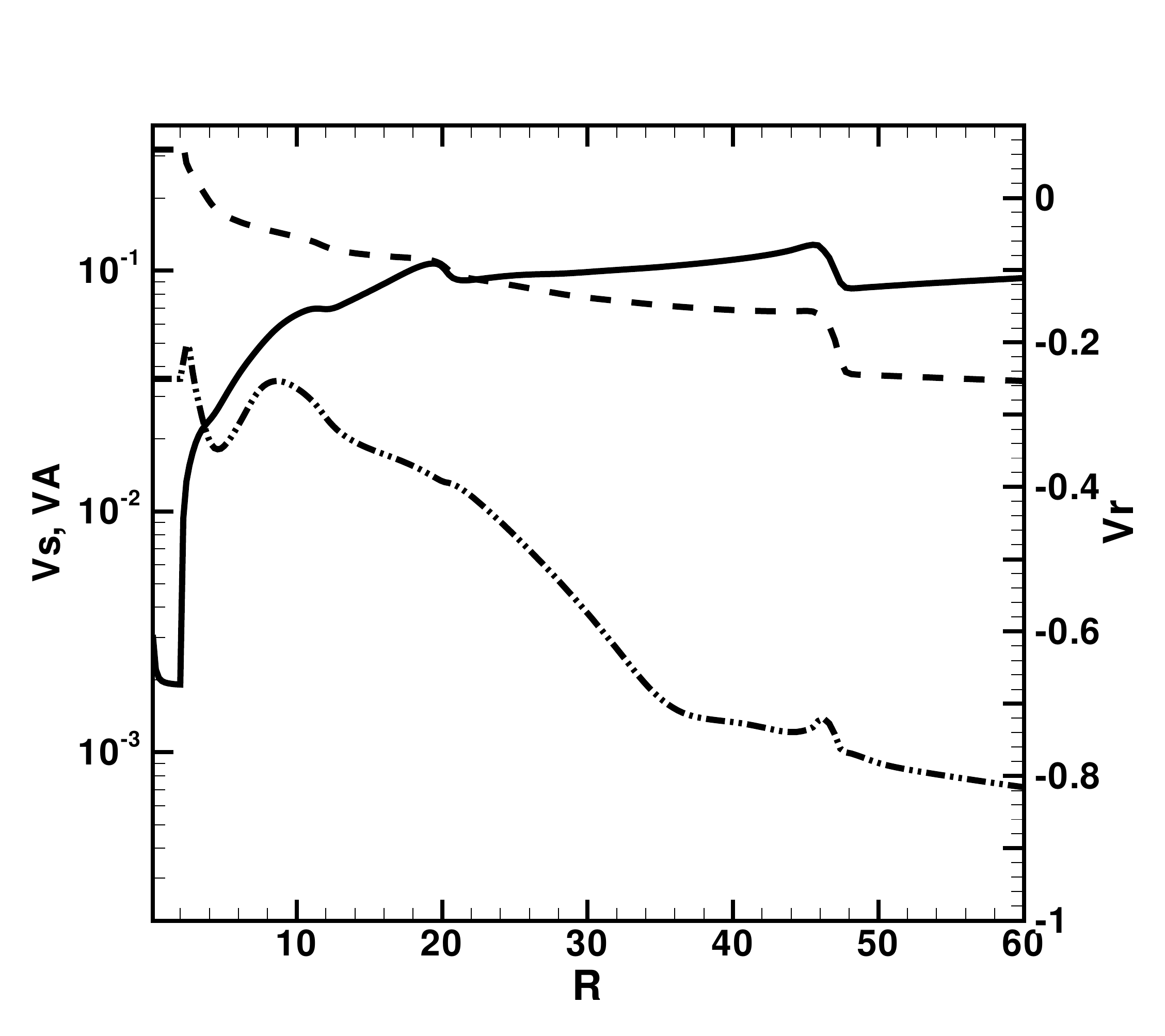}
        \label{fig:8a}
      \end{minipage}

      \begin{minipage}{0.45\linewidth}
        \centering
        \includegraphics[width=0.8\textwidth]{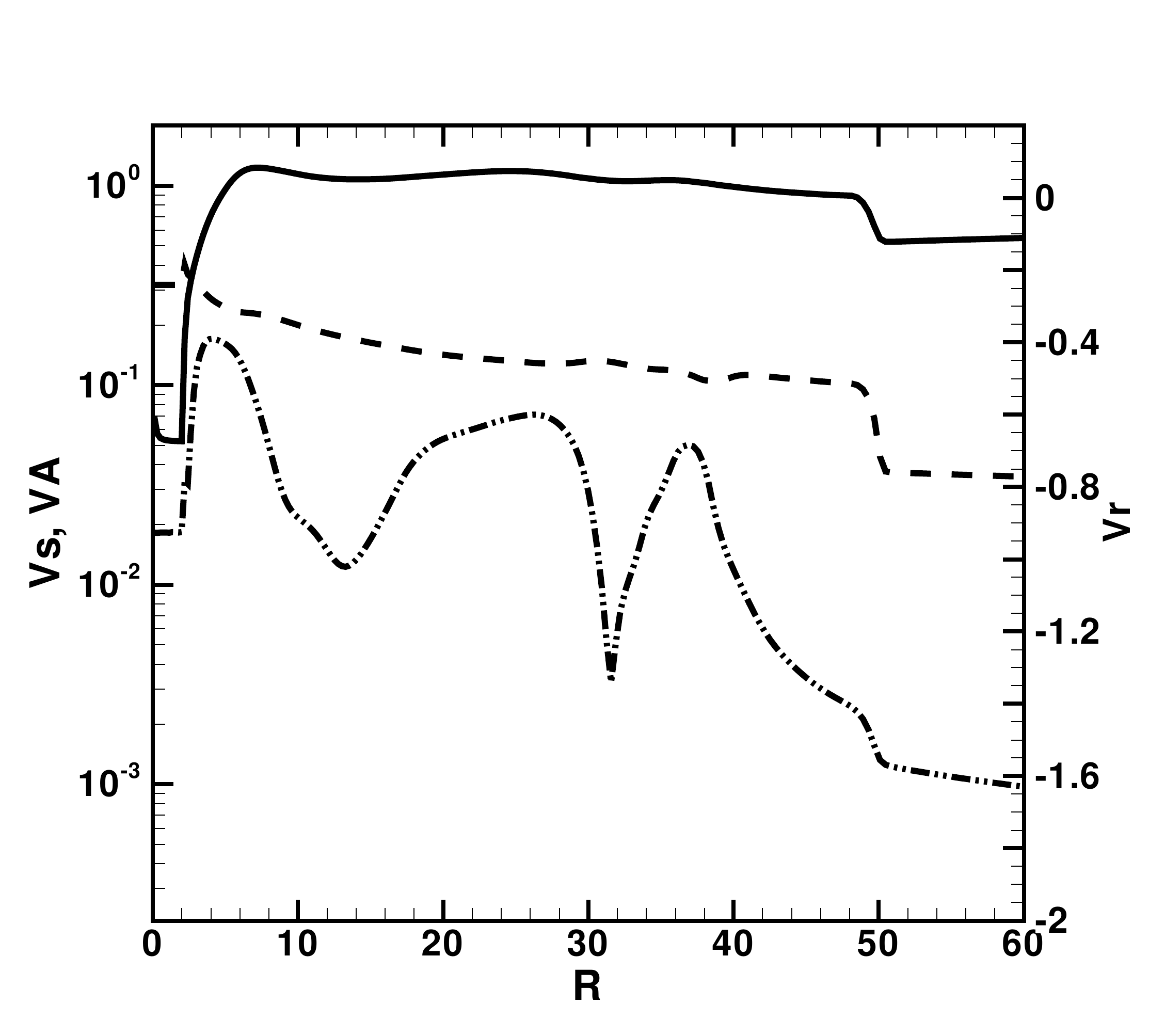}
        \label{fig:8b}
      \end{minipage} \\

    \end{tabular}
\caption {Profiles of radial velocity $V_{\rm r}$ (solid line), sound velocity $V_{\rm s}$ (dashed line) and Alfv\'{e}n velocity $V_{\rm A}$ (dash-dot-dot line) on the equator at $t = 1.936 \times 10^7$ s (left-hand panel) and  $t = 1.980 \times 10^7$ (right-hand panel). 
 The oscillating shock locations are found at $R \sim$ 48 on the shock descending branch (left-hand panel) and $\sim$ 50 on the shock ascending branch (right-hand panel). 
In the latter case,  the outflow triggered by the strong magnetic field near the event horizon is expanding as a hot blob at $R \sim 40$ and the dip at $R \sim$ 30 of the Alfv\'{e}n velocity
 is due to the weak spot of magnetic field behind the expanding hot blob.}
  \label{fig:7}
  \end{figure*}

\subsubsection{Time delay between 22,  43 and 350 GHz flares}
 Time delay of $\sim$ 20 min between flares at 22 and 43 GHz
 have been detected  with simultaneous X-ray observations for Sgr A*. To examine the 
time delay between the radio bands from the shock oscillation model, we recalculated 2D  radiation MHD simulations during
$t$= 1.9237 -- 2.0 $\times 10^7$ s,  using further smaller time step of 20$R_{\rm g}/c$
($\sim$ 13 min).
 Figs.~8 and ~9 show  two cases of monochromatic luminosity curves of  $L_{\rm 22}$ at 22 GHz (blue square), $L_{\rm 43}$ at 43 GHz (orange Delta), $L_{\rm 350}$ at 350 GHz (black diamond) and $R_{\rm s}$ (green circle) during $t$= 1.925-- 1.960 $\times 10^7$ s and  1.966-- 1.986 $\times 10^7$ s, respectively,  over a  shock oscillation cycle.
 The ``arrow" shows local peaks or dips on the luminosity curves where 
 A1-- 12, B1 -- B12 and C1-- C12 are the corresponding event names in $L_{\rm 22}$, $L_{\rm 43}$ and  $L_{\rm 350}$ curves, respectively.  The number within the bracket of B and C shows the delay point number relative to the 22 GHz flare. 
Here, one point time interval is $20 R_{\rm g}/c$  ($\sim$ 13 min). 
The scale of $L_{\rm 350}$ is shown inside on the right vertical axis.
 $L_{\rm 22}$ delays $L_{\rm 43}$ by $\sim$ 1 -- 2 points time interval (13 -- 26 min)
on the shock ascending branch  and conversely  $L_{\rm 43}$ delays $L_{\rm 22}$ by 
1 -- 4 points time interval except for event A1 and A8 on the shock descending branch. 
$L_{\rm 350}$ delays  $L_{\rm 22}$ variously by $\sim$ 1 -- 11 points time interval (13 min -- 2.5 h) positively on the shock descending branch and negatively on the shock ascending branch, respectively.

\begin{figure*}
    \begin{tabular}{cc}

      \begin{minipage}{0.7\linewidth}
        \centering
        \includegraphics[width=100mm,height=80mm]{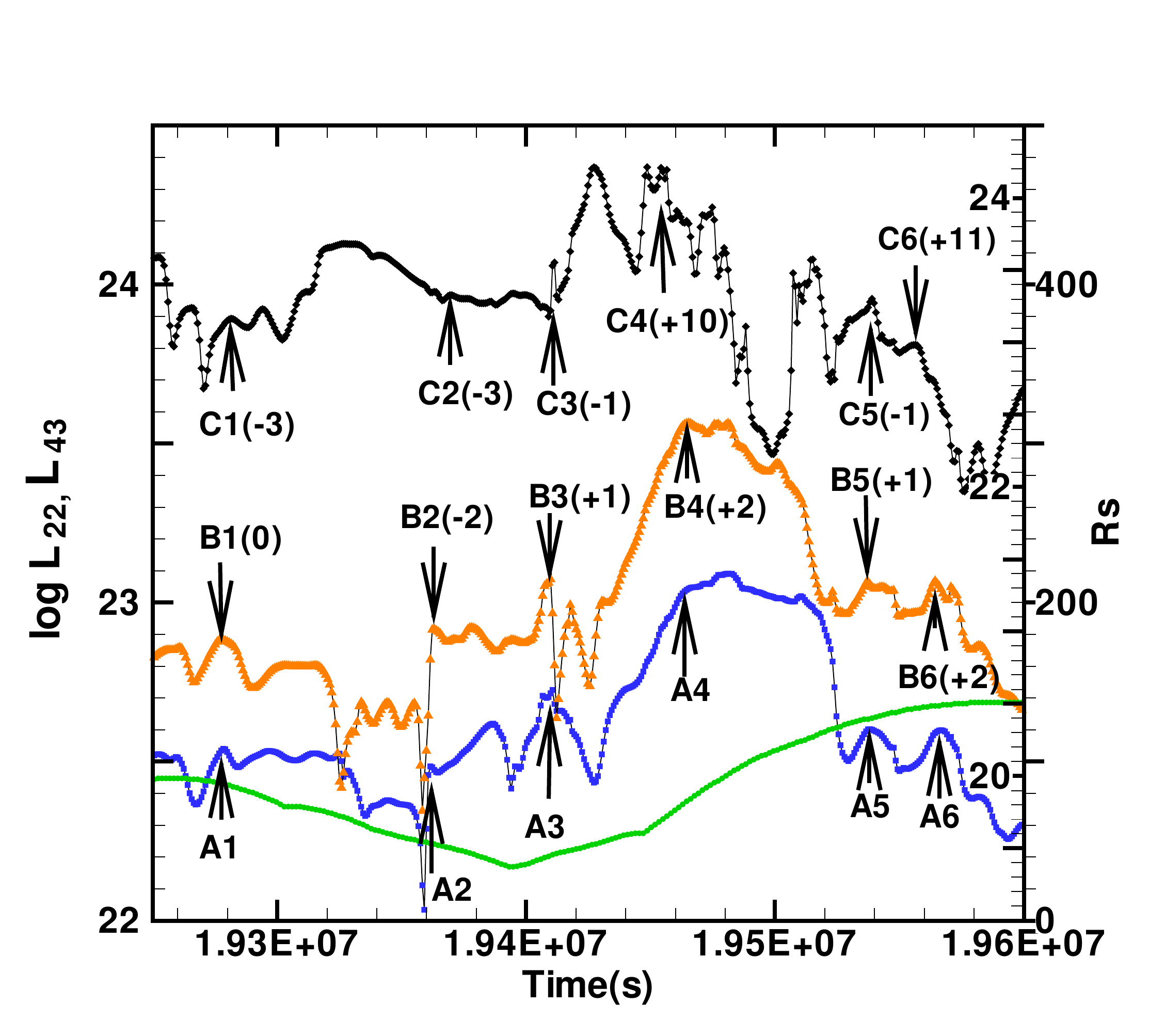}
        \label{fig:8}
      \end{minipage}

      \begin{minipage}{0.0\linewidth}
        \centering
        \includegraphics[width=0.8\textwidth]{fig8.pdf}
        \label{fig:8}
      \end{minipage} \\
    \end{tabular}
\caption{Monochromatic  luminosity curves of  $L_{\rm 22}$ (erg ${\rm s}^{-1} {\rm Hz}^{-1}$) at 22 GHz (blue square), $L_{\rm 43}$ (erg ${\rm s}^{-1} {\rm Hz}^{-1}$) at 43 GHz (orange Delta) , $L_{\rm 350}$ (erg ${\rm s}^{-1} {\rm Hz}^{-1}$) at 350 GHz (black diamond) and oscillating shock location $R_{\rm s}$ (green circle) on the equator during $t$= 1.925-- 1.960 $\times 10^7$ s in model Rad1. The ``arrow" shows local peaks or dips on the luminosity curves where  A1-- A6, B1 -- B6 and C1-- C6 are the corresponding event names in $L_{\rm 22}$, $L_{\rm 43}$, and  $L_{\rm 350}$ curves, respectively, the number within the bracket of B and C shows the delay point  number relative to $L_{\rm 22}$ and negative number means time delay to 22 GHz flare.  The scale of $L_{\rm 350}$ is shown inside on the right vertical axis. Here, one point time interval corresponds to $20 R_{\rm g}/c$ ($\sim$ 13 min).  }
  \label{Fig: fig8}
  \end{figure*}

\begin{figure*}
    \begin{tabular}{cc}

      \begin{minipage}{0.7\linewidth}
        \centering
        \includegraphics[width=0.8\textwidth]{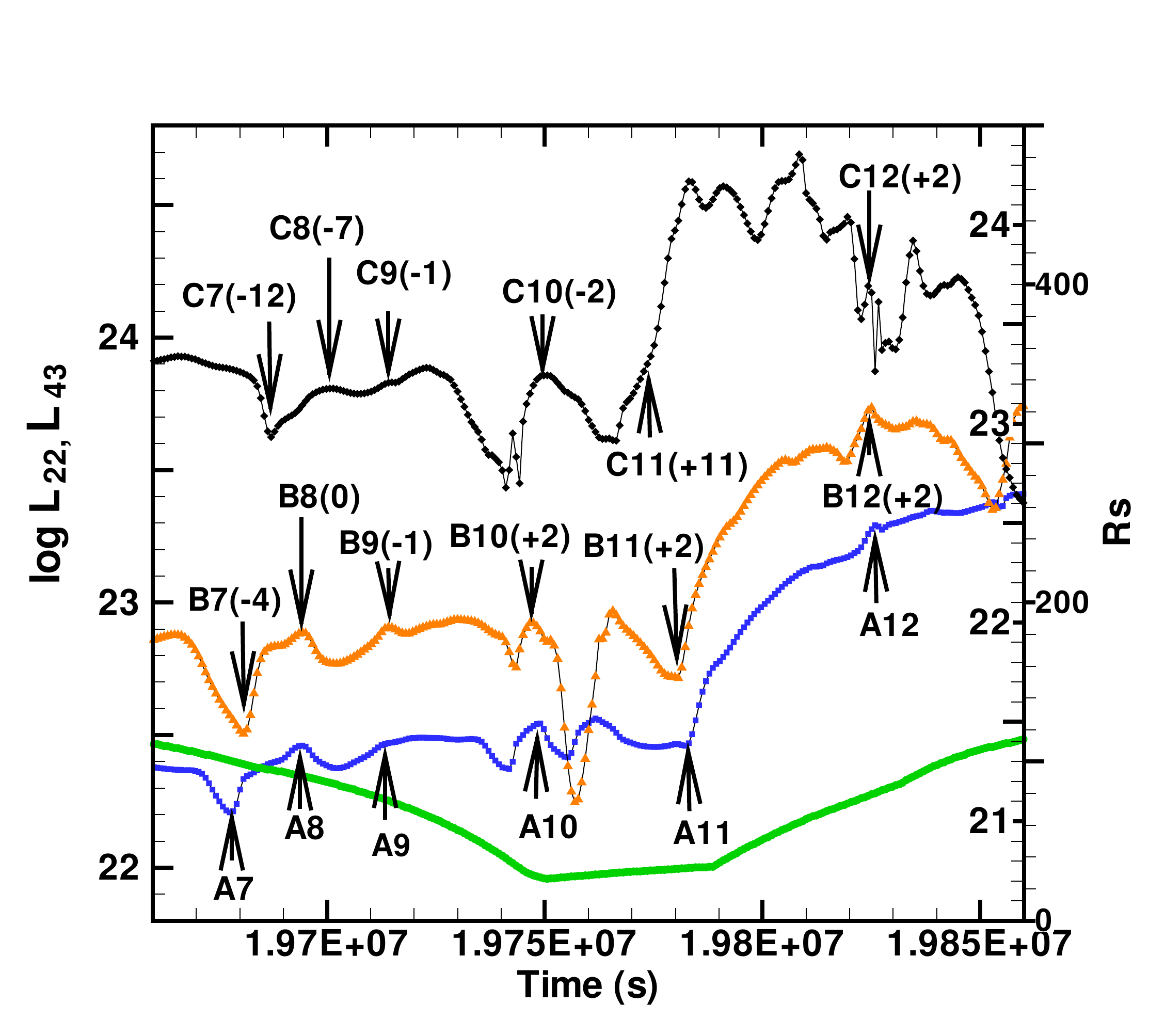}
        \label{fig:9}
      \end{minipage}

      \begin{minipage}{0.0\linewidth}
        \centering
        \includegraphics[width=0.8\textwidth]{fig9.pdf}
        \label{fig:9}
      \end{minipage} \\
    \end{tabular}
\caption{Same as Fig.~8 but for events A7 -- A12, B7 -- B12 and C7-- C12 during  $t$= 1.966 -- 1.986 $\times 10^7$ s.
  }
  \label{Fig: fig9}
  \end{figure*}

\begin{figure*}
    \begin{tabular}{cc}
      \begin{minipage}{0.45\linewidth}
        \centering
        \includegraphics[width=0.9\textwidth]{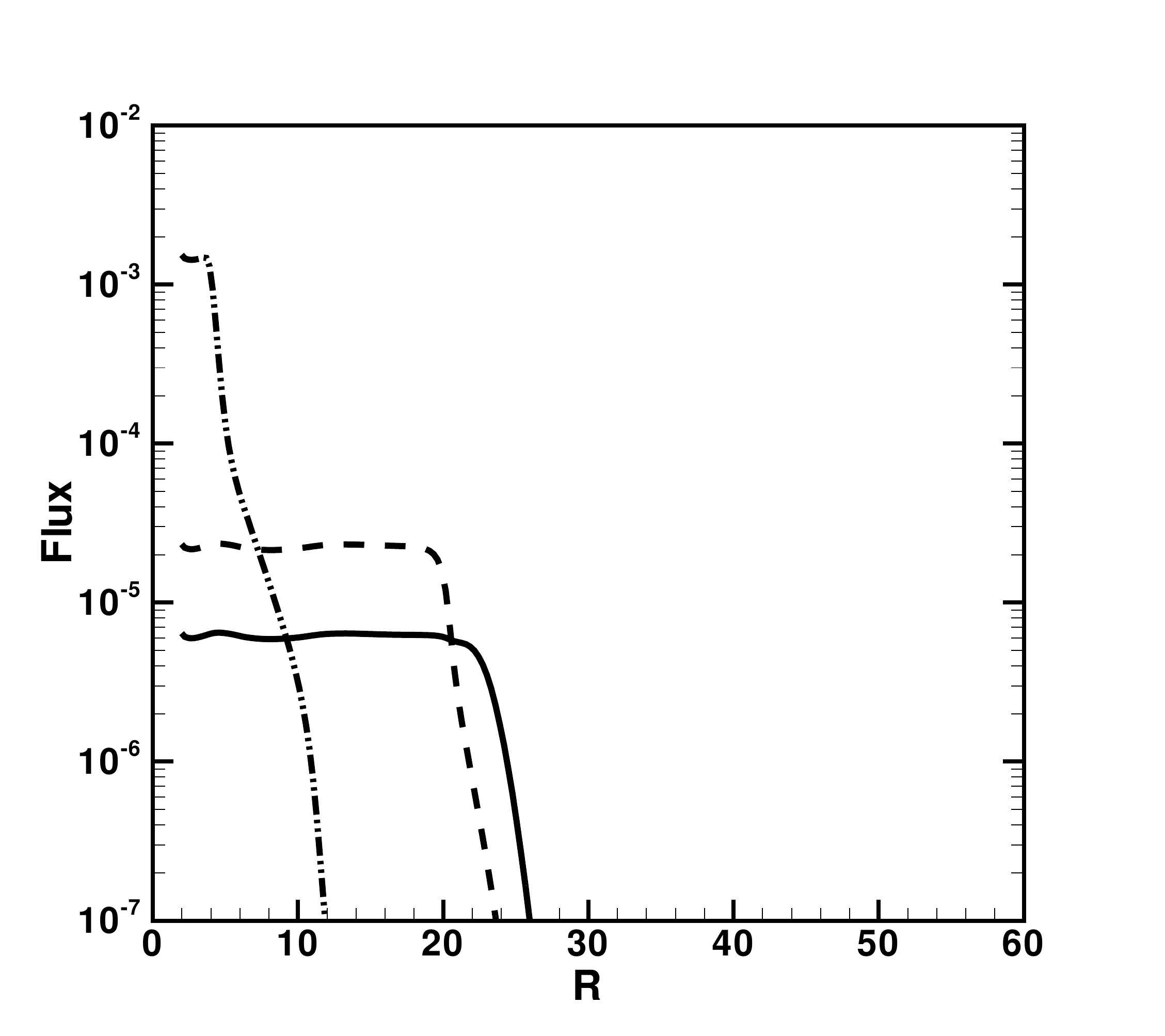}
        \label{fig:10a}
      \end{minipage}

      \begin{minipage}{0.45\linewidth}
        \centering
        \includegraphics[width=0.9\textwidth]{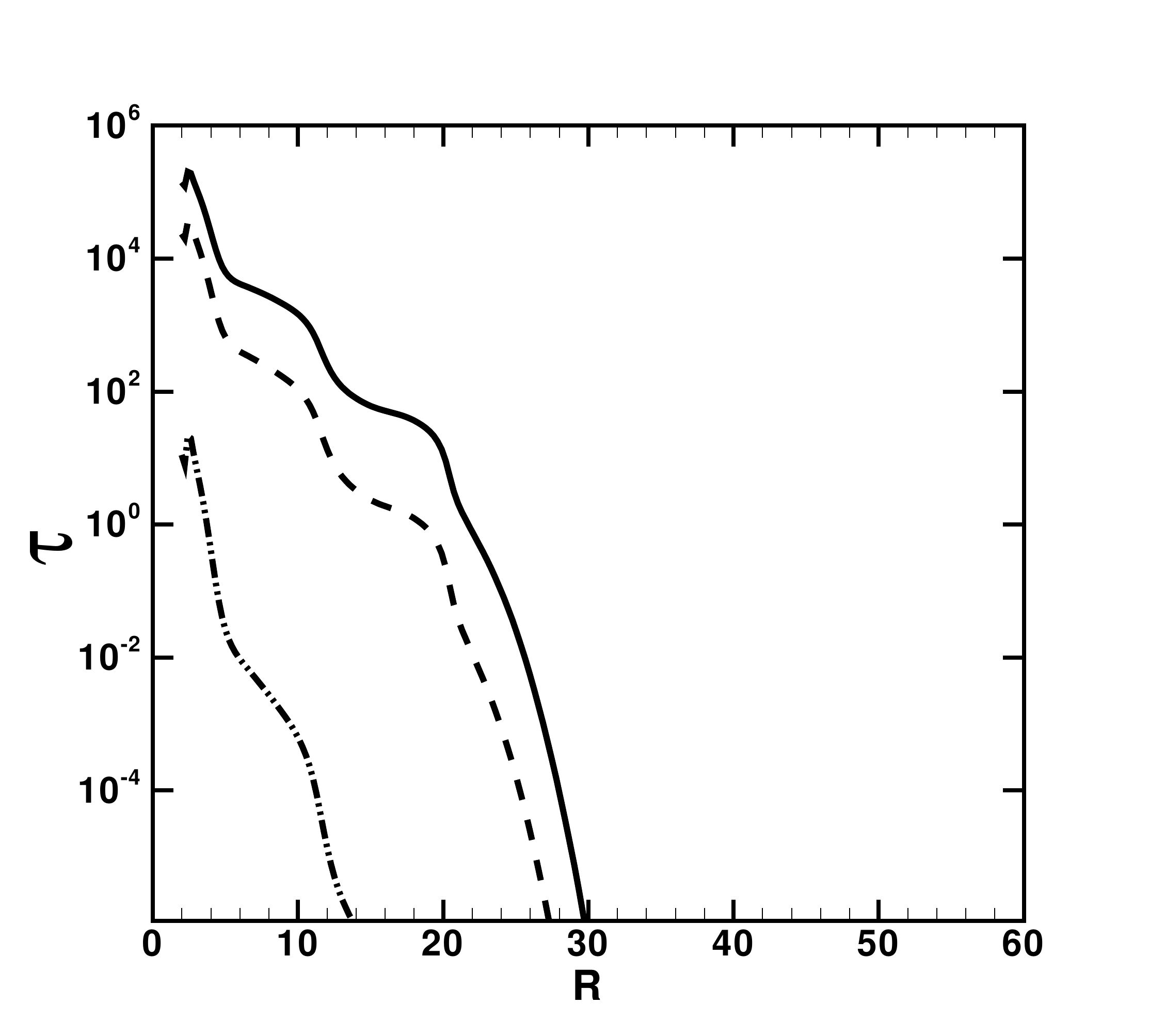}
        \label{fig:10b}
      \end{minipage} \\
    \end{tabular}
\caption {Monochromatic radiation flux (erg ${\rm s}^{-1}{\rm cm}^{-2} {\rm Hz}^{-1}$) (left-hand panel) and optical thickness (right-hand panel) at the disc surface in the frequency of
 22 GHz (solid line), 43 GHz (dashed line) and 350 GHz (dash-dot-dot line) at  $t = 1.936 \times 10^7$ s for model Rad1.
}
  \label{fig:10}
  \end{figure*}

\begin{figure*}
    \begin{tabular}{cc}
      \begin{minipage}{0.45\linewidth}
        \centering
        \includegraphics[width=0.9\textwidth]{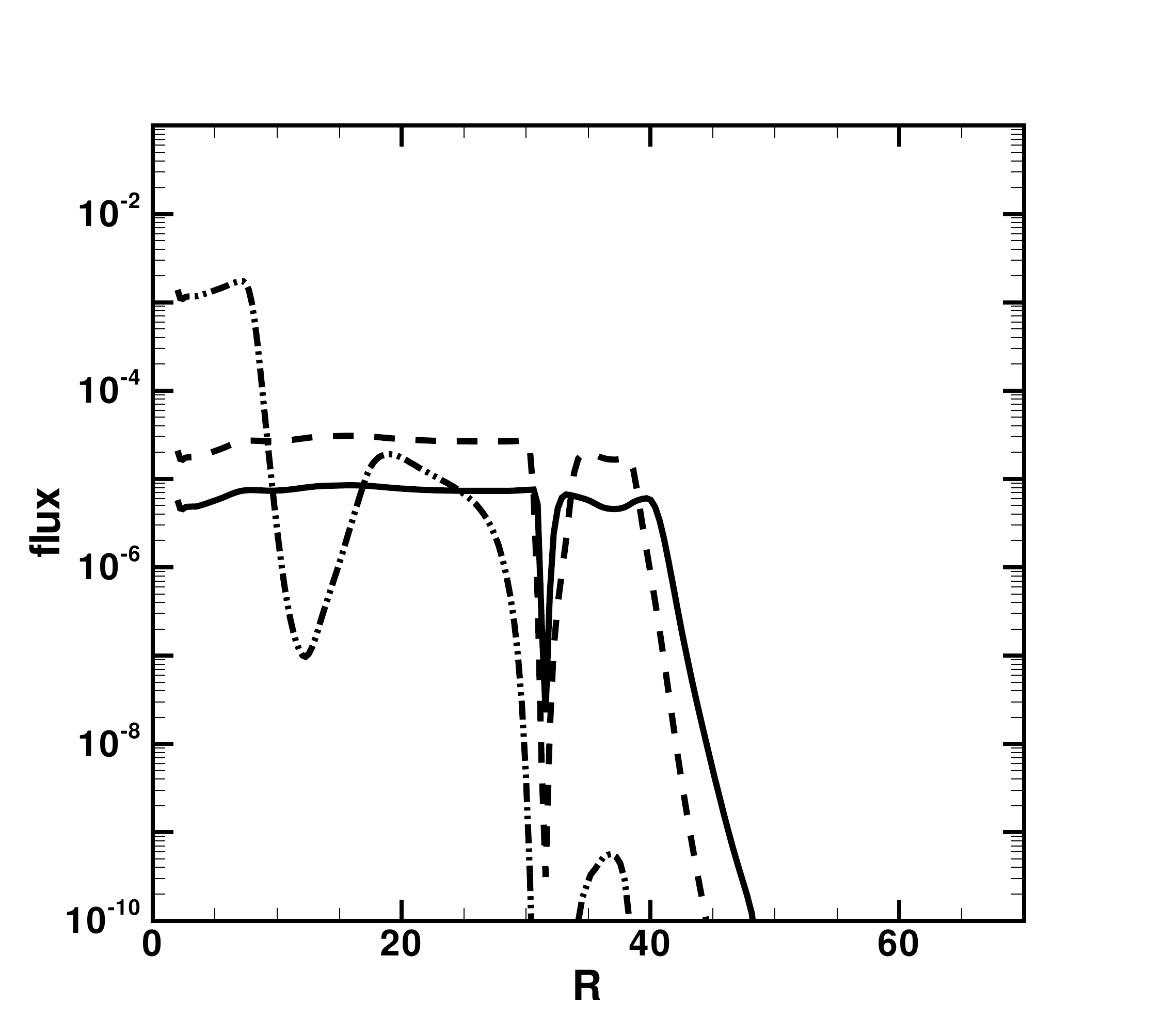}
        \label{fig:11a}
      \end{minipage}

      \begin{minipage}{0.45\linewidth}
        \centering
        \includegraphics[width=0.9\textwidth]{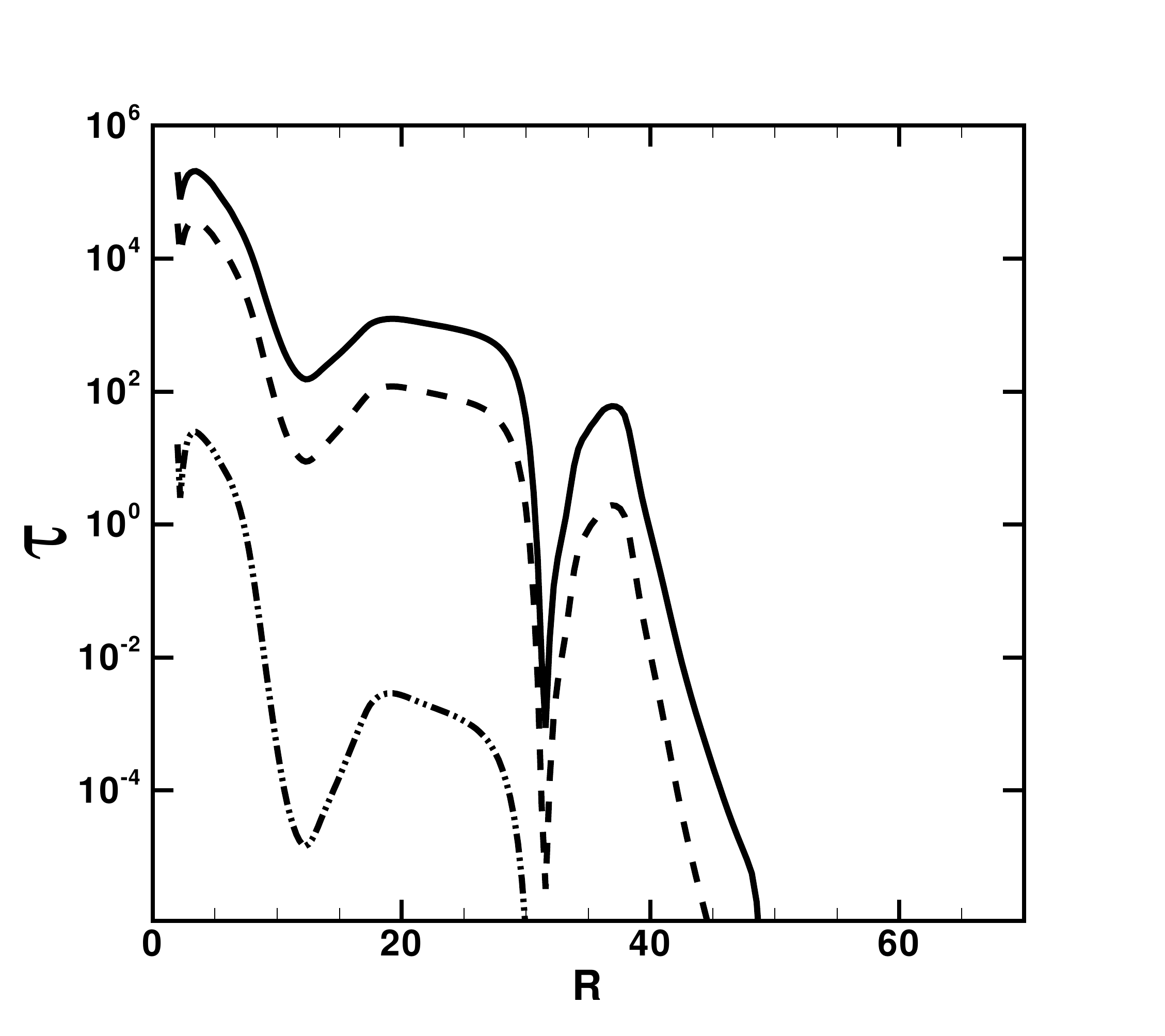}
        \label{fig:11b}
      \end{minipage} \\
    \end{tabular}
\caption {Same as Fig.~10 but for $t = 1.980 \times 10^7$ s. }
  \label{fig:11}
  \end{figure*}

 These time delays are understood from consideration of the frequency-dependent effective radii
 of $R_{\rm 22}$, $R_{\rm 43}$ and $R_{\rm 350}$, at whicht the optical thickness $\tau _{\rm R, \nu}$ at frequency $\nu$ becomes $\sim$ unity. Figs.~10 and ~11 show the radial dependence of vertical 
radiation flux $F_{\rm R, \nu}$  and optical thickness $\tau_{\rm R, \nu}$ at the accretion disc surface
for $t = 1.936 \times 10^7$  and 1.980 $\times 10^7$ s, which correspond
 to event A2 in Fig.~8 and event R6 in Fig.~6, respectively.
We find here that the radiation fluxes are negligibly small beyond the frequency-dependent effective radii $R_{\rm 22}$, $R_{\rm 43}$ and $R_{\rm 350}$.
That is, most of the monochromatic luminosities are produced within the frequency-dependent effective  radii.  $R_{\rm 22}$, $R_{\rm 43}$ and $R_{\rm 350}$ are $\sim$ 22.0, 20.0  and 4.0 at $t = 1.936 \times 10^7$ s and 31.0, 30.0 and 7.0 at $t$ = 1.980 $\times 10^7$ s, respectively, from Figs.~10 and ~11.
 These effective radii are dependent only on $\kappa_{\rm \nu}$, that is, the density and the temperature, but not the magnetic field.
 Then, the delay times between 22 and 43 GHz flares and between 22 and 350 GHz flares are estimated
 as the transit time of the sound velocity,  ($R_{\rm 22}$ - $R_{\rm 43})/V_{\rm s}$ and  ($R_{\rm 22}$ - $R_{\rm 350})/V_{\rm s}$, respectively, which are roughly 20 and 180$R_{\rm g}/c$ ($\sim$ 13 min and 2 h) with $V_{\rm s} = 0.1c$ at $t = 1.936 \times 10^7$, and 10 and 240$R_{\rm g}/c$ ($\sim$ 8 min and 2.5 h) with $V_{\rm s} \sim  0.1c$
at  $t = 1.980 \times 10^7$ from the velocity profiles in Fig.~7.
Figs.~10 and 11 also show the existence of the optically thick region in the radio frequencies. 
These optically thick regions appear at  22 -- 350 GHz emissions considerd here only in the inner region of the accretion flow.

The predicted delay time between 22 and 43 GHz flares agree well the numerical ones.
However, the delay times between very different frequencies such as 22 and 350 GHz seem to differ from the actual numerical delay time.
 This may be partly related to the broader optically thick region  between $R_{\rm 22}$ and $R_{\rm 350}$
 because the transit time of the light passing through the optically thick region is not specified definitly.
 While, in the transit region between $R_{\rm 22}$ and $R_{\rm 43}$, the optically thick gas is confined in very
 narrow region and the predicted delay time may agree with the numerical ones.
 
\subsubsection{Time delay between 8  and 10 GHz flares}
 
  Remarking on  the recently   detected simultaneous observation at 8 and 10 GHz with VLA
( Michail, Yusef-Zadeh \& Wardle 2021), we examined the time delay between 8 and 10 GHz flares.
  Fig.~12 shows the monochromatic luminosity curves  $L_{\rm 8}$ at 8 GHz (blue square),  $L_{\rm 10}$ at
 10 GHz (orange circle) and the oscillating shock location $R_{\rm s}$ (green circle)  on the equator during $t$= 1.940 -- 1.950 $\times 10^7$ s in the shock oscillation model.
 The ``arrow" shows local peaks on the luminosity curves where 
 A1-- A5 and  B1 -- B5 are the corresponding event names in $L_{\rm 8}$ and  $L_{\rm 10}$, respectively, the number in the bracket of 10 GHz event B shows the delay point number relative to $L_{\rm 8}$. Although the flares show marginal one point delay time  except for 5 points delay time in event B3, we consider them to be significant from comparison and analogy with the time delay in 23 -- 43 GHz flare and conclude
 the delay time between 8 and 10 GHz flare as 1 point time interval ( $\sim$ 13 min). 
The 8 GHz flare  delays 10 GHz flare on the shock ascending branch but precedes 10 GHz flare on the shock descending branch as well as the time-delay relation between 22 and 43 GHz flares.
For instance, taking the event A2 at $t = 1.9412 \times 10^7$, the effective radii $R_{\rm 8}$ at 8 GHz 
and $R_{\rm 10}$ at 10 GHz are found to be 35.5 and 33.0, respectively. Then, the predicted delay time  between
8 and 10 GHz flare is 25  $R_{\rm g}/c \sim$ 16 min with $V_{\rm s} \sim 0.1c$ which agrees well with the numerical delay time of $\sim$ 13 min.

\begin{figure*}
    \begin{tabular}{cc}

      \begin{minipage}{0.8\linewidth}
        \centering
        \includegraphics[width=100mm,height=80mm]{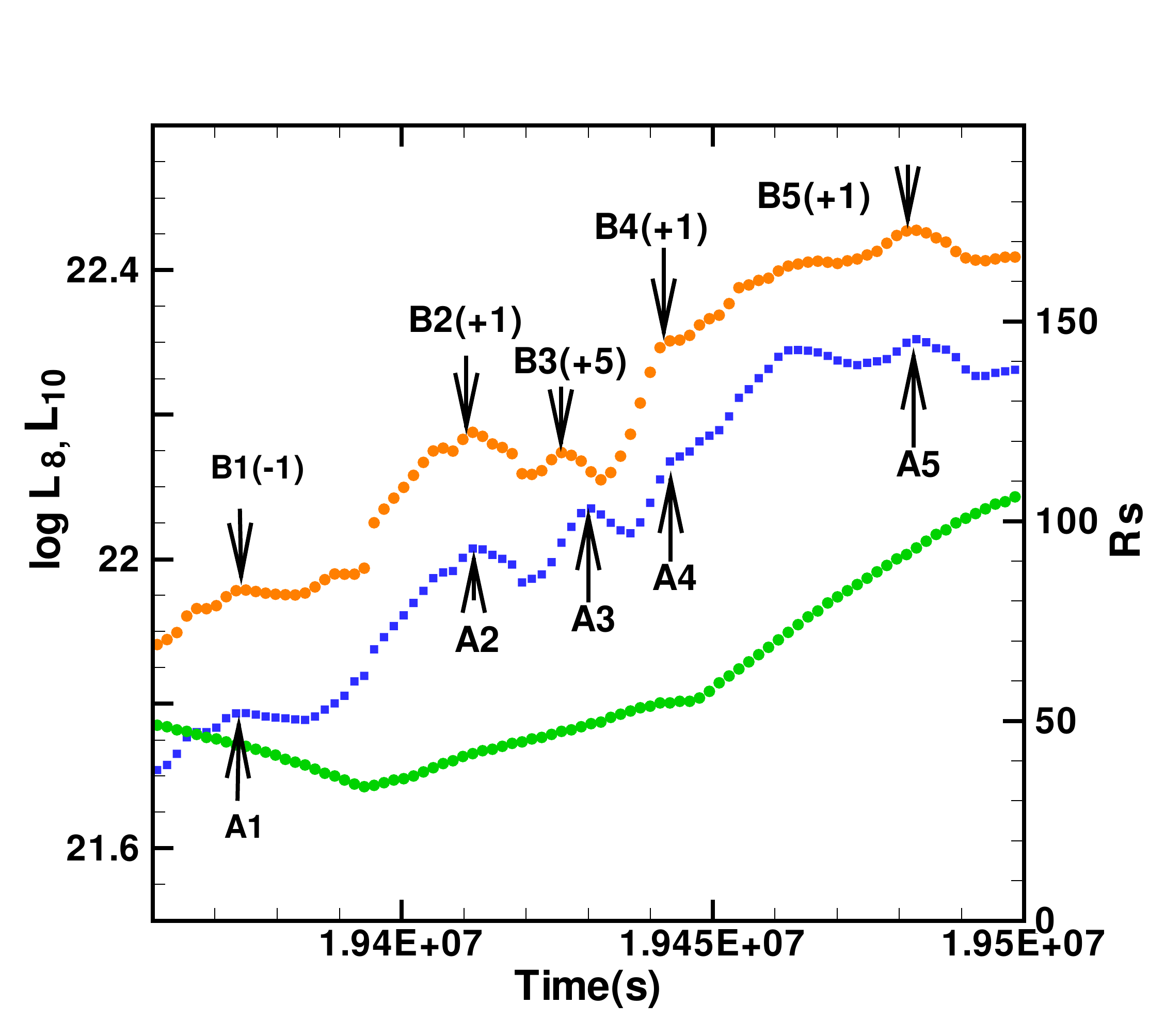}
        \label{fig:12}
      \end{minipage}

      \begin{minipage}{0.0\linewidth}
        \centering
        \includegraphics[width=0.8\textwidth]{fig12.pdf}
        \label{fig:12}
      \end{minipage} \\
    \end{tabular}
\caption{Monochromatic  luminosity curves of  $L_{\rm 8}$  (erg ${\rm s}^{-1} {\rm Hz}^{-1}$) at 8 GHz (blue square), $L_{\rm 10}$ (erg ${\rm s}^{-1} {\rm Hz}^{-1}$) at 10 GHz (orange circle) and oscillating shock
 location $R_{\rm s}$ (green circle) on the equator  during $t$= 1.940-- 1.950 $\times 10^7$ s in model Rad1. The ``arrow" shows local peaks on the luminosity curves where 
 A1-- A5 and B1 -- B5 are the corresponding event names in $L_{\rm 8}$ and $L_{\rm 10}$ curves, respectively. The number in the bracket of 10 GHz flare B shows the delay point number relative to $L_{\rm 8}$.
 The delay time between 8 and 10 GHz flare is 1 point time interval ( $\sim$ 13 min) except 5 points interval
in event B3. The 8 GHz flare delays 10 GHz flare on the shock ascending branch but precedes 10 GHz flare on the shock descending branch as well as the flare relation between 22 and 43 GHz.
 Here one point time interval corresponds to $20 R_{\rm g}/c$  ($\sim$ 13 min).
  }
  \label{Fig: fig12}
  \end{figure*}

\begin{figure}
\begin{center}
        \includegraphics[width=0.5\textwidth]{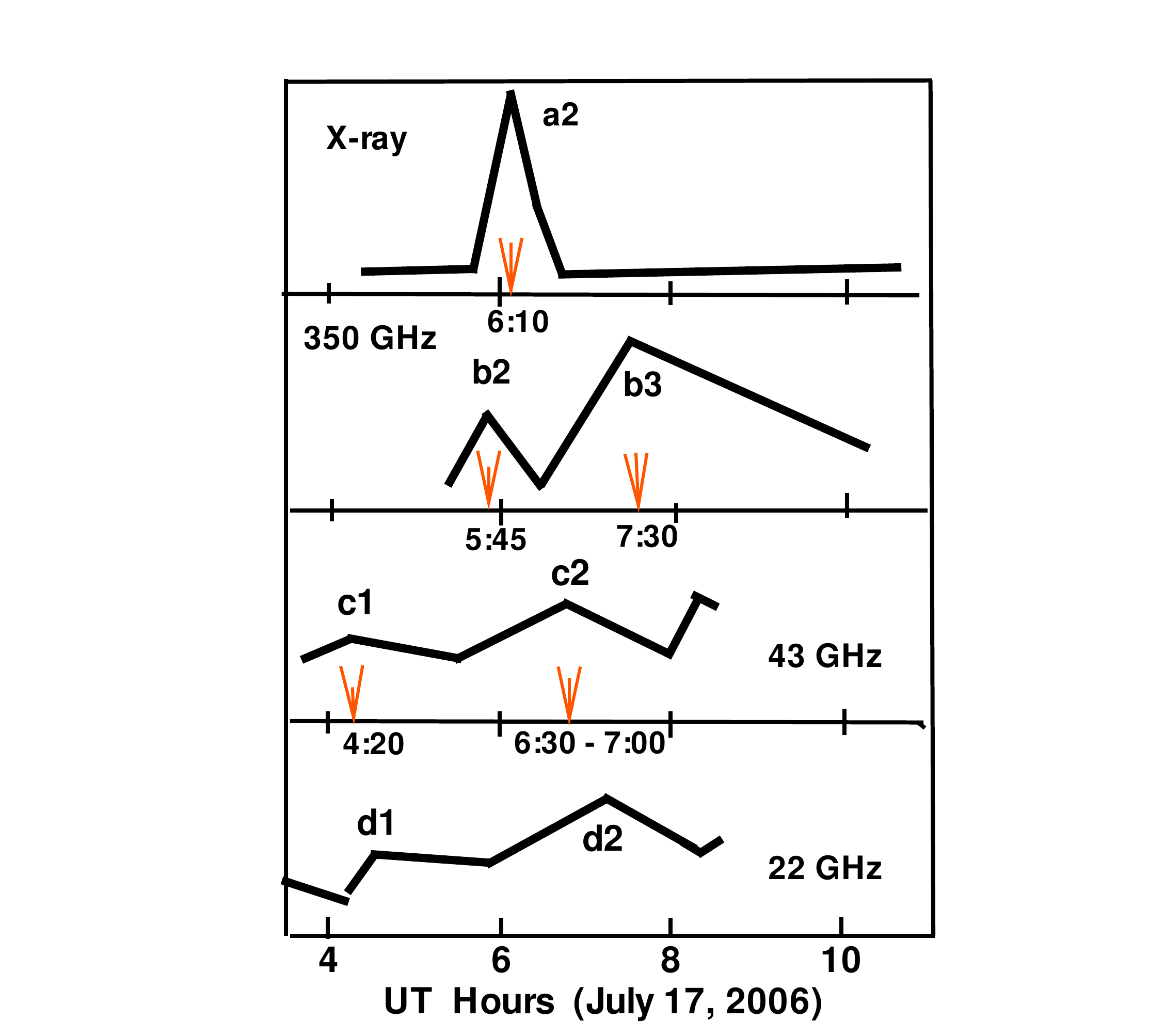}
\caption {Schematic light curves of X-ray, 350 GHz, 43 GHz and 23 GHz bands on 2006 July 17 for Sgr A*
 taken from Fig.~1 in the paper by Yusef-Zadeh et al. (2008). a2, b2, b3, c1, c2, d1 and d2 show flares at
 each energy bands. Only horizontal axis of time is roughly shown for time delay discussion. Arrow shows 
 rough flare peak time.  We suggest that a series of flares  a2, b2. c2 and d2 may be associated each other.
}
\end{center}
\end{figure}

\begin{figure}
\begin{center}
        \includegraphics[width=0.5\textwidth]{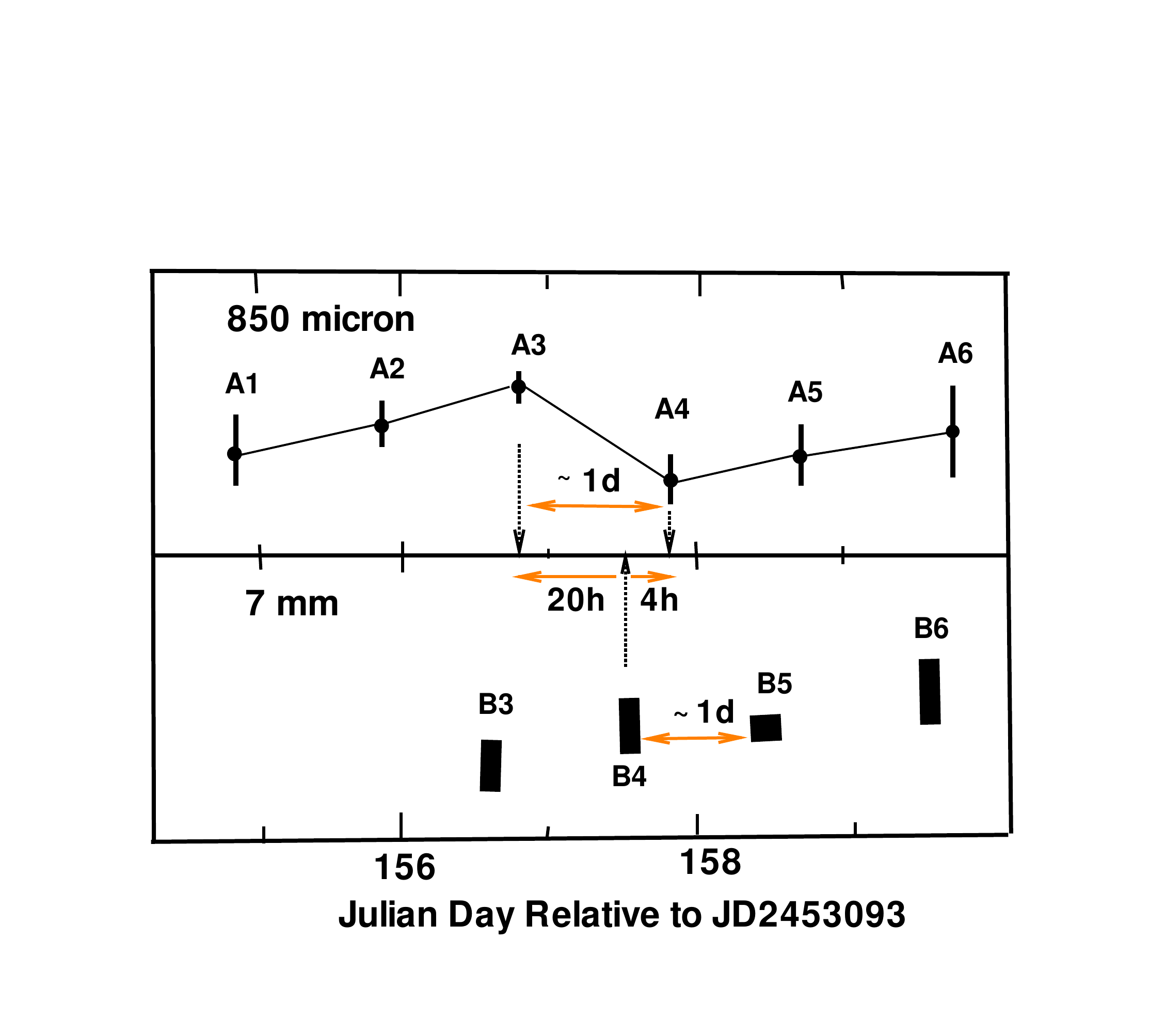}
\caption {Schematic light curves of 850 $\mu m$ (350 GHz) and 7 mm at the second epoch on 2004 September
 for Sgr A*
 taken from Fig.~8 in the paper by Yusef-Zadeh et al. (2006a), where  A1 -- A6 and B3 -- B6 show  flares 
at 850 $\mu m$ and 7 mm in radio bands, respectively.
  Only horizontal axis of time is roughly shown for time delay discussion and the flares appear roughly  with a period of 1d. 
}
\end{center}
\end{figure}

\subsubsection{Comparison with the Plasmon model }
The expanding hot plasmon model (van der Laan 1966) has been proposed to explain the time delay of 20 -- 40 min between the flare peaks observed at 22 and 43 GHz in several different epochs using the VLA (Yusef-Zadeh et al. 2006b, 2008).
Following van der Laan (1966), the synchrotron flux from a homogeneous 
blob can be written as
\begin{equation}
    S_\nu(R) = S_{0} \left(\frac{\nu}{\nu_0}\right)^{2.5}
    \left(\frac{R}{R_0}\right)^3 \,
    \frac{1-\exp(-\tau_{\rm \nu})}{1-\exp(-\tau_0)}
    \label{eq:Snu},
\end{equation}
where $R(t)$ is the radius of the expanding hot blob and $R_{\rm 0}$ is the initial blob radius, and the optical depth 
\begin{equation}
    \tau_{\rm \nu} = \tau_0 \left(\frac{\nu}{\nu_0}\right)^{-(K+4)/2}
    \left(\frac{R}{R_0}\right)^{-(2K+3)}.
    \label{eq:tau}
\end{equation}
Here  $K$ is the index of the relativistic particle 
energy spectrum as $n(E)\propto E^{-K}$, $\tau_0$ is the critical optical depth at the maximum of the 
light curve at any frequency, which satisfies 
\begin{equation}
    e^{\tau_0}-(2K/3+1)\tau_0 - 1 = 0
    \label{eq:tau0}
\end{equation}
and $\nu_0$ is the frequency at which this occurs when $R=R_0$. 
This equation is derived from the condition $\partial_t S(\nu,t)=0$ and $\tau_0$ ranges 0.55 to 1.9 as K ranges from 0.5 to 3.

From the above relations, the initially  optically thick ($\tau_{\rm \nu} \gg 1$) synchrotron flux  $S_\nu(R)$
 at a fixed frequncy $\nu$ increases with increasing radius as $\propto R^3$ but turns to decrease with further 
 increasing radius as $\propto R^{-2K}$ when the optical thickness becomes small as $\tau_{\rm \nu} \ll 1$.
 If we consider the synchrotron flares at frequencies $\nu_1$ and  $\nu_2$ ($\nu_1 < \nu_2$) and the frequency-dependent radii of $R_{\rm \nu_1}$ and $R_{\rm \nu_2}$ at which the optical thickness is $\tau_0$,
  $R_{\rm \nu_1}$ must be larger than $R_{\rm \nu_2}$ from the relation of equation (20). 
Accordingly, the low frequency flare (the maximum flux at the low frequency) appears after the high frequency flare (the maximum flux at the high frequency)  because the hot blob is expanding.
In this respet, our analytical method for the time delay prediction on the shock ascending
 branch is identical to the idea of the plasmon model because the radiation flux 
 has same functional form of the optical depth in both cases.
We compare the frequency-dependent radii of $R_{\rm \nu_{22}}$, $R_{\rm \nu_{43}}$ and $R_{\rm \nu_{350}}$ at
 22, 43 and 350 GHz by the plasmon model with the numerical ones in the shock oscillation model.
The frequency-dependent radius $R_{\rm \nu}$ is given by, from equation (20)
\begin{equation}
    R_{\rm \nu}  = \left(\nu \over \nu_0\right)^{-(K+4) \over {2(2K+3)}} \;R_0.
    \label{eqRnu}
\end{equation}
As the result,  $R_{\rm \nu_{22}}$, $R_{\rm \nu_{43}}$ and $R_{\rm \nu_{350}}$ are 8.0, 6.2 and 2.7
 with K=3, $\nu_0$=130 GHz and $R_0$ = 4, and 19.0, 13.0 and 4.0 with K=0.5, $\nu_0$ = 350 GHz
 and $R_0$=4, respectively (see Fig.~3 in Yuan-Zadeh et al. 2006b and Fig.~6b in Yusef-Zadeh et al. 2008).
 These values of $R_{\rm \nu_{22}}$, $R_{\rm \nu_{43}}$ and $R_{\rm \nu_{350}}$ are 
 somewhat small compared with the frequency-dependent
 effective radii of 22.0, 20.0 and 4.0 at $t = 1.936\times 10^7$ and 31.0, 30.0 and 7.0 at $t = 1.98\times 10^7$ s in our simulations.
 This may be partly due to that the frequency-dependent radius in our model is the radius at which the optical depth is unity but not $\tau_0$.
 However, the remarkable difference between the plasmon model and the shock  oscillation
 model is that the former consider only the expanding hot blob, while in the latter  the 
 time delay of flares between the radio and  X-ray and between  the narrow radio bands
 appear on the shock ascending branch as well as the plasmon model but  occurs inversely also on the shock descending branch.

\section{Comparison with observations of Sgr A*}
 We compare the time lag between the synchrotron and bremsstrahlung emissions in the 
shock oscillation model with  that between radio and X-ray emission observed in Sgr A*. 
Through the simultaneous radio, near-infrared (NIR), and X-ray observation campaign with  HST NICMOS, XMM-Newton and VLA,  the observations have shown that radio flares delays X-ray flares by $\sim$ 2 h on  2006 July (Yusef-Zadeh et al. 2008),  
 up to $\sim$ 5 h on 2007 April  (Yusef-Zadeh et al.  2009),  $\le$ 80 min on 2013 July, $\sim 2$ h  on 2013 September,  $\le$ 7.5 h or $\le$ 3.9 h on 2013 October and  $\sim$ 30 min on 2014 May (Capellupo et al. 2017), $\sim$ 30 min on 2019 July (Michail et al. 2021), and  20 -- 30 min on 2019 July (Boyce et al. 2022). Most of the delay time is 0.5 -- 5 h
 and agrees roughly with the time delay of 2 -- 3 h obtained from the numerical simulations.
 However, our simulations predict that radio flare delays X-ray flare on the shock descending branch
 with inflow but X-ray delays radio flare on the ascending branch with outflow.

 Ponti et al. (2015, 2017) found a very bright flare from Sgr A*, 
which is by more than two orders of magnitude higher than the bremsstrahlung emission
in usual flares, starts in NIR and then an X-ray flare follows after $\sim 10^3$ s.
They confirm the origin of the very bright X-ray flare as the synchrotron nature. 
If the NIR flare is simultaneous with radio flare by the original synchrotron emission, the very bright flare may appear as Type C as is found in event R6 in Fig.~6. 
 The time delay is comparable with the time delay of $\le$ 0.55 h predicted from the shock oscillation model.

 Yusef-Zadeh et al. (2008) report that  22 GHz flare delays 43 GHz by 20 -- 30 min.  
 The delay time is comparable to 13 -- 26 min in our simulations.
  Michail, Yusef-Zadeh \& Wardle (2021) report the detection and analysis of radio flares
 across narrow  16 frequency bands between 8 and 10 GHz on 2014 April and find 
that the flare delays successively with 
 decreasing frequencies and 8 GHz light curve delays 10 GHz light curve by 18 min, which
 agrees well with $\sim$13 min obtained from the present simulations.
 The positive or negative time delay relation is dependent on 
 the shock location branch as well as the case of  22 and 43 GHz flare.

A composite light curve of flaring activity in the X-ray, 350 GHz, 43 GHz and 22 GHz on 2006 July 
 reported by Yusef-Zadeh et al. (2008) is very interesting because we are able to interpret the observational result with our time delay analysis from the simulations. 
Fig.~13 shows schematic light curves of X-ray, 350 GHz, 43 GHz and 23 GHz bands for Sgr A*
 taken from Fig.~1 in the paper by Yusef-Zadeh et al. (2008). Here, a2, b2, b3, c1, c2, d1 and d2 show the flare
 events at each energy bands. Only horizontal axis of time is roughly shown for the time delay interpretation. 
A strong X-ray flare (a2) and weak and strong 350 GHz flares (b2 and b3)  are detected at 
$\sim$ 6:10, 5:45 and  7:30 hr UT on 2006 July. Two flares c1 and c2 at 43 GHz  occur at $\sim$ 
4:25 and 6:30 -- 7:00 hr UT, respectively and
 22 GHz flares (d1, d2) delay the 43 GHz flare (c1, c2) by $\sim$ 20 -- 30 min. 
 From the plasmon model and our flare analysis on the shock ascending branch,
  22, 43 and 350 GHz flares should successively delay in this order. Then we may recognize the flares b2, c2 and d2 as a relevant group. The strong flare 350 GHz (b3) perhaps correspond to  next relevant group at 22 and 43 GHz flare although it may be ambiguous at the final phase of the light curve.
Then, the 43 GHz flare (c2) delays the 350 GHz flare (b2) by $\sim$ 45 -- 75 min.
Such delay time between 43 and 350 GHz is observed  as $\sim$ 100 min at events A4 
and A11 in Figs.~8 and ~9 of the simulations.  If  X-ray flare (a2) at the top panel in 
 Fig.~13 is associated with the radio flare of 350 GHz flare (b2), the time delay of $\sim$ 25 min between X-ray and radio flare is reasonable compared with $\sim$ 30 min in  events R4 and R7 in Figs.~5 and ~6 on the ascending branch in the shock oscillation model.  
Finally, we suggest that a series of flares  a2, b2. c2 and d2 may be associated each other, which appears as Type C flare.

The plasmon model predicts that the low frequency flare always delays the high frequency flare,
 However, there exist an important observation of flares at radio bands, where the high frequency flare delays the low frequency flare.
  Yusef-Zadeh et al. (2006a) showed  simultaneous X-ray, 850 $\mu$m (0.85 mm),
  450 $\mu$m, 3mm and 7 mm flare light curves at two epochs on 2004 March -- September of Sgr A*. 
 Fig.~14 shows schematic light curves of 850 $\mu m$ (350 GHz) and 7 mm at the second epoch
 taken from Fig.~8 in the paper by Yusef-Zadeh et al. (2006a), where  A1 -- A6 and
 B3 -- B6 show  flare points at 850 $\mu m$ and 7 mm, respectively.
  Here, consecutive four pairs of flare (A3, B3) -- (A6, B6) with a period of $\sim$ 1d  are regarded to correlate to each
 other simultaneously and  850 $\mu m$ flares delay 7 mm flares roughly by $\sim$ 4 h.
 Although another pairs of (A3, B4) and (A4, B5) may be considered to associate together, the delay time of $\sim$ 20 h is too long to explain the time delay from the analysis of the shock oscillation model. 
Similarly, from Fig.~7 in Yusef-Zadeh et al. (2006a), we suggest that the successive three 850 $\mu m$ flares delay 7 mm flars about by $\sim$ 1 h.
These positive delays of the high frequency to the low frequency flare contradict the plasmon model but are naturally expected in the shock oscillation model.

\section{Summary and Discussion}
We examined the time delay relations of flares  between radio and X-ray and between
 narrow radio frequency bands in Sgr A*, from analyses of the synchrotron and bremsstrahlung luminosity and monochromatic luminosity  $L_{\rm \nu}$ curves obtained from the time-dependent 2D relativistic radiation MHD simulations based on the shock oscillation model.
  Here, ion and electron temperatures are obtained by  solving the radiation energy equilibrium equation  that Coulomb collisions transfer energy from ions to electrons equals to the sum of synchrotron and bremsstrahlung cooling rates in electrons.
 The results can be summarized as follows:

(1) We find  three types of time delay between the synchrotron and bremsstrahlung emissions which correspond to radio and X-ray emissions, respectively;
 Type A with a time delay of  2 - 3 h, Type B with no time delay and Type C with an inverse negative time delay of  $\le$ 0.5 -- 1 h.
 Type A flare occurs on the shock  descending branch, Type B appears when the shock is far away from the center and Type C  occurs only when the oscillating shock contracts up to the minimum location, 
 an expanding hot blob triggered by the sporadic magnetic field is incorporated into the  oscillating shock, and as a result a strong flare is formed.
 
(2) The synchrotron emission is always dominant in a core region of 5$R_{\rm g}$ size, while most of bremsstrahlung emission mainly originates in a region of 10 -- 20$R_{\rm g}$ at the minimum shock location but comes from a distant region of $\le 50R_{\rm g}$  at the maximum shock location.
  The time delay of Type A and Type C  is interpreted as the transit time of the Alfv\'{e}n and acoustic waves
  between the synchrotron and  bremsstrahlung dominant regions, respectively. 
 The delay time of Type A is $\sim 250R_{\rm g}/c \sim$ 2.5 h
 with the Alfv\'{e}n velocity $\sim 0.02c$ but in Type B  there is actually no interaction between the
 synchrotron and bremsstrahlung dominant regions because the transit time is too long.
 On the other hand, the delay time of Type C is small as $\sim$ 0.5 h because the sound velocity
 far exceeds the Alfv\'{e}n velocity.

(3)   22 GHz flare delays 43 GHz flare by $\sim$ 13 -- 26 min on the shock ascending branch
 and conversely precedes 43 GHz flare  by the same delay time on the shock descending branch.
 The delay time between 22 and 350 GHz ranges extensively up to 4 h.
  The delay time between 8 and 10 GHz flare is $\sim$ 13 min as well as the time delay
 relation between 22 and 43 GHz.
 These time delays between 22 and 43 GHz and  between 8 and 10 GHz agree well
 with 20 -- 30 min and 18 min, respectively, observed in Sgr A*.
 The time delay of flares between the narrow radio frequency bands is interpreted  as the
 transit time of the acoustic wave between the frequency-dependent effective radii 
 $R_{\rm \tau_{\rm \nu}=1}$ at which the optical depth $\tau_{\rm \nu}$ is unity. 

(4)  Many delay type between radio and X-ray flares observed in Sgr A*  belongs to Type A 
  with time lags of $\sim$ 0.5 -- 5 h  and Type C is rarely observed with time lag of $\le$ 20 min, 
 as is found on 2006 July 17 (Yusef-Zadeh et al. 2008) and  2014 August 30 (Ponti et al. 2015, 2017).
 Similarly, it is likely that the flare at the higher radio frequency delays that at the lower radio frequency by $\sim$ 
10 -- 30 min and also the inverse delay occurs, as is confirmed in the observed flares on  2004 September
 (Yusef-Zadeh et al. 2006a) and on 2014 April 17 (Michail, Yusef-Zadeh \& Wardle 2021), respectively.

The shock oscillation model for the long-term flare of Sgr A* explains well the observations 
of the flaring rate and also the delay time of flares between radio and X-ray
 and between the narrow frequency radio bands.
Of course, the energy spectra of Sgr A*  do not consist of only the simple
 synchrotron and bremsstrahlung emissions and are reproduced by various  radiation mechanisms
 such as the synchrotron, the inverse Compton and the synchrotron self-Compton
 emissions including a hybrid electron population consisting of both thermal and nonthermal particles (e.g. Yuan, Quataert \& Narayan 2003, 2004). However, the present estimate of 
 time delay between the radio and X-ray emissions and between the narrow radio bands  may not be altered  largely.

 The expanding hot plasmon model
 explains not only the time delay between the peak emission of 43 and 22 GHz  but also other available measurements of the flux values in near-IR, submillmeter, milimeter, and radio wavelengths, using the adquate parameters fitted to the observed light curves.
  To explain the submillimeter light curves correlated with the NIR and possible delays of $\sim$ 30 min
 between 22 and 43 GHz , Witzel et al. (2021) proposed a simple source model with three-step-process 
 of electron injection with a nonthermal energy distribution into a spherical central region, compression of magnetic field lines and increasing magnetic flux, and adiabatic expansion with no injection of electrons,
 and modeled the three-step process as a cyclic process in a single zone.
 The model is similar to the plasmon model but differs from that it considers the cyclic 
expansion and contraction processes of the blob.

In the shock oscillation model, the monochromatic radiation energy flux 
$F_{\rm \nu}(R,\tau_{\rm \nu})$ on the shock descending branch begins to brighten abruptly 
when the optical depth $\tau_{\rm \nu}$ becomes unity.
Since $\tau_{\rm \nu} \propto \kappa_{\rm \nu} \propto 1/\nu^2$ for radio waves,
 a larger optical depth $\tau_{\rm \nu}$ at a smaller frequency $\nu$ leads to a  larger  effective radius $R_{\rm \tau_{\rm \nu} =1}$
because the optical depth decreases with increasing radius.
This means that a brightening of radiation flux $F_{\rm \nu}(R,\tau_{\rm \nu})$ begins progressively 
from a flare with the smallest frequency at inflow phase of the accretion flow. 
Contrarily, when the intermittently increasing magnetic field near the event horizon yields to an outflow
 and expanding hot blob as Type C flare, the extinction of  the brightened optically thick flux starts in
a  flare with  the largest frequency and ends with the smallest frequency flare which delays most.
Therefore, the oscillating shock model predicts that, when radio flare delays X-ray flare, 43 GHz flare  
 necessarily delays 22 GHz flare, and vice versa.
  In section 4, we already confirmed an existence of observation for Sgr A* such that the high frequency
 flare delays the low frequency flare, which contradicts the plasmon model but is natural in the shock oscillation
 model.
Another observational constraint on the shock oscillation model is that the positive delay of 
radio to X-ray flare have to appear together with the positive delay of high frequency to low frequency flare,
 if these flares occur simultaneously.
At present,  such confirmatory observation on the oscillating shock model has not been met yet, 
 including denial observation.
  We hope future simultaneous observations in radio to X-ray and in narrow radio bands  for Sgr A*
  confirm them.

\section*{Acknowledgments}
CBS is supported by the National Natural Science Foundation of China under grant no. 12073021. The numerical computations were conducted on the Yunnan University Astronomy Supercomputer.

\section*{Data Availability}
The numerically created data that support the findings of this study are available from the corresponding authors upon reasonable request.

\label{lastpage}

\end{document}